\RequirePackage{fix-cm}
\documentclass[twocolumn]{svjour3}          
\smartqed  
\usepackage{graphicx}
\usepackage{mathptmx}      
\usepackage{latexsym}
\usepackage{graphicx}
\usepackage{subfigure}
\usepackage{amsmath}
\usepackage{amssymb}
\usepackage{latexsym}
 \usepackage{cancel}
 \usepackage{multirow}
 \usepackage{color}
 \usepackage{epstopdf}%
\def \bn {\mathbf{n}}
\def \tr {\tilde{r}}
\def \Hc {H_{\infty}^{\star}}
\begin{document}

\title{Viscous Taylor droplets in axisymmetric and planar tubes: from Bretherton's theory to empirical models
}


\author{Gioele Balestra         \and
        	    Lailai Zhu		\and
	    Fran\c{c}ois Gallaire 
}


\institute{Gioele Balestra \at
              Laboratory of Fluid Mechanics and Instabilities, EPFL, CH1015 Lausanne, Switzerland \\
              Tel.: +41 21 69 35312\\
              Fax: +41 21 69 37855\\
              \email{gioele.balestra@epfl.ch}           
           \and
           Lailai Zhu \at
              Laboratory of Fluid Mechanics and Instabilities, EPFL, CH1015 Lausanne, Switzerland \\
              Linn\'{e} Flow Centre and Swedish e-Science Research Centre 
(SeRC), KTH Mechanics, SE 10044 Stockholm, Sweden\\
              Department of Mechanical and Aerospace Engineering, Princeton 
University, NJ 08544, USA \\
              \and
              Fran\c{c}ois Gallaire \at
              Laboratory of Fluid Mechanics and Instabilities, EPFL, CH1015 Lausanne, Switzerland
}

\date{Received: date / Accepted: date}

\maketitle

\begin{abstract}
The aim of this study is to derive accurate models 
for quantities characterizing the dynamics of droplets of non-vanishing 
viscosity in capillaries. In particular, we propose models for the uniform-film 
thickness separating the droplet from the tube walls, for the droplet front and 
rear curvatures and pressure jumps, and for the droplet velocity in a range of 
capillary numbers, $Ca$, from $10^{-4}$ to $1$ and inner-to-outer viscosity 
ratios, $\lambda$, from $0$, \textit{i.e.}\ a bubble, to high viscosity droplets. Theoretical asymptotic results obtained in 
the limit of small capillary number are combined with accurate numerical 
simulations at larger $Ca$. With these models at hand, we can compute 
the pressure drop induced by the droplet. 
The film thickness at low capillary numbers ($Ca<10^{-3}$) agrees well with Bretherton's scaling for bubbles as long as $\lambda<1$. For larger viscosity ratios, the film thickness 
increases monotonically, before saturating for $\lambda>10^3$ to a value $2^{2/3}$ times larger than the film thickness of a bubble. At larger capillary numbers, the film thickness 
follows the rational function proposed by Aussillous \& Qu\'er\'e 
\cite{aussillous2000quick} for bubbles, with a fitting coefficient which is 
viscosity-ratio dependent. This coefficient modifies the value to which the film 
thickness saturates at large capillary numbers. The velocity of the droplet is 
found to be strongly dependent on the capillary number and  viscosity 
ratio. We also show that the normal viscous stresses at the front and rear caps 
of the droplets cannot be neglected when calculating the pressure drop for 
$Ca>10^{-3}$.
\end{abstract}

\section*{List of symbols}
\noindent
\noindent
\begin{tabular}{ll}
  $A$ & coefficient for flow profile \\
  $B$ & coefficient for flow profile \\
  $C$ & coefficient for interface profile of static meniscus \\
  $\mathcal{C}$ & mean curvature of droplet interface  \\
  $D$ & coefficient for interface profile of static meniscus \\
  $c_1$, $c_2$ & coefficient for fitting law of $P$, $\bar{P}$ \\
  $Ca$ & capillary number based on droplet velocity \\
  $Ca_{\infty}$ & capillary number based on mean outer velocity \\
  $F$ & coefficient for minimum film thickness \\
  $\bar{F}$ & averaged $F$ coefficient \\
  $G$ & coefficient for minimum film thickness \\
  $H$ & thickness of film between wall and droplet \\
  $H_{\textsf{min}}$ & minimum film thickness \\
  $H_{\infty}$ & uniform film thickness \\
  $H_{\infty}^{\star}$ & critical uniform film thickness for recirculations\\
  $K$ & coefficient for linearized lubrication equation \\
  $\tens{I}$ & identity tensor\\
  $L_d$ & droplet length \\
  $M$ & coefficient for pressure model \\
  $m$ & rescaled viscosity ratio \\
  $N$ & coefficient for pressure model \\
  $\vec{n}$ & unit vector normal to the droplet interface \\
  $O$ & coefficient for pressure model \\
\end{tabular}

\noindent
\noindent
\begin{tabular}{ll}
  $P$ & coefficient for interface profile of static meniscus \\
  $\bar{P}$ & averaged $P$ coefficient \\
  $p$ & pressure \\
  $p_{\textsf{linear}}$ & pressure if constant gradient \\
  $Q$ & coefficient for uniform film thickness model \\
  $q$ & volume flux \\
  $R$ & capillary tube radius or half width \\
  $Re$ & Reynolds number \\
  $r$ & radial direction (axisymmetric geometry) \\
  $\tilde{r}$ & half width of droplet \\
  $S$ & coefficient for classical pressure model \\
  $t$ & time \\
  $T$ & coefficient for plane curvature model \\
  $U_{d}$ & droplet velocity \\
  $U_{\infty}$ & average outer flow velocity \\
  $u_{\infty}$ & outer far-field velocity profile \\
  $\vec{u}$ & velocity field \\
  $u$ & streamwise velocity \\
  $v$ & spanwise velocity \\
  $x$ & streamwise direction (planar geometry) \\
  $y$ & spanwise direction (planar geometry) \\
  $z$ & axial direction (axisymmetric geometry) \\
  $Z$ & coefficient for plane curvature model \\
\end{tabular}

\section*{Greek symbols}
\noindent
\begin{tabular}{ll}
  $\alpha$ & parameter for solution of linear lubrication equation \\
  $\beta$ & coefficient for plane curvature model \\
  $\Delta$ & difference between inner and outer quantities \\
  $\Delta p^{\text{NP}}$ & pressure correction due to non-parallel flow effects\\
  $\Delta p_{\textsf{tot}}$ & total pressure drop\\
  $\gamma$ & surface tension \\
  $\eta$ & rescaled film thickness \\
  $\kappa$ & plane curvature of droplet interface in ($z,r$) or ($x,y$)  \\
  $\kappa_{f,r}$ & plane curvature at the front/rear droplet extremities \\
  $\lambda$ & inner-to-outer dynamic viscosity ratio  \\
  $\mu$ & dynamic viscosity \\
  $\xi$ & rescaled axial direction \\
  $\tens{\sigma}$ & total stress tensor \\
  $\tens{\tau}$ & viscous stress tensor \\
  $\phi$ & phase of solution of linear lubrication equation \\
  $\chi$ & geometric coefficient \\
  $\Omega$ & droplet volume or area \\
\end{tabular}\\

\section*{Subscripts and superscripts}
\noindent
\begin{tabular}{ll}
  $f$ & front cap \\
  $i$ & inner \\
  $o$ & outer \\
  $r$ & rear cap \\
  $zz$ & normal tensor component in the axial direction
\end{tabular}\\

\section*{Abbreviations}
\noindent
\begin{tabular}{ll}
  2D & two-dimensional \\
  3D & three-dimensional \\
  ALE & arbitrary Lagrangian-Eulerian \\
  BIM & boundary integral method \\
  FEM & finite element method \\
\end{tabular}\\

\section{Introduction \label{tx:introduction}}

Two-phase flows in microfluidic devices gained considerably in importance during the last two decades 
\cite{stone2004engineering,gunther2006multiphase}. The key for success of these microfluidic tools 
is the fluid compartmentalization, allowing the miniaturization and manipulation 
of small 
liquid portions at high throughput rates with a limited number of necessary 
controls. Reduced liquid quantities are commonly used as individual reactors in 
several biological and chemical applications \cite{kohler2014micro}, as well as 
in industrial processes \cite{abadie2012hydrodynamics} and in micro-scale heat 
and mass transfer equipments 
\cite{leung2011effect,warnier2010pressure,mikaelian2015bubbly}.
Bubbles and droplets often flow in microchannels with a round or 
rectangular/square cross-section 
\cite{kreutzer2005inertial,khodaparast2015dynamics,mikaelian2015bubbly}.

\begin{figure}[ht!]
\includegraphics[width=0.45\textwidth]{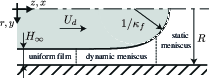}
\caption{Sketch of the front meniscus of the bubble advancing at velocity $U_d$ in a capillary of radius $R$ with indication of the uniform thin-film region of thickness $H_{\infty}$, the dynamic meniscus region and the static meniscus region. The plane curvature $\kappa_f$ of the front static cap in the ($z,r$) or ($x,y$) plane is also highlighted. 
\label{fig:geometryZones}}
\end{figure}

The dynamics of a bubble in a microchannel has been the subjects 
of several studies, 
since the seminal works of Fairbrother \& Stubbs \cite{fairbrother1935119}, 
Taylor \cite{taylor1961deposition} and Bretherton \cite{bretherton1961motion}. 
These long bubbles, also referred to as \textit{Taylor bubbles}, flowing in a 
tube of radius $R$, have been characterized by the thickness $H_{\infty}$ of the 
uniform film separating them from the tube walls, the minimum thickness $H_{\textsf{min}}$ of the film, the plane curvature of the 
front and rear caps in the ($z,r$) or ($x,y$) plane, $\kappa_f$ and $\kappa_r$, as well as by their velocity 
$U_d$. Bretherton \cite{bretherton1961motion} used a lubrication approach to 
derive the asymptotic scalings in the limit of small capillary numbers, $Ca = 
\mu_o U_d/\gamma$, where $\mu_o$ is the dynamic viscosity of the outer fluid 
and $\gamma$ the surface tension. In particular, Bretherton 
\cite{bretherton1961motion} showed that in the limit of $Ca \rightarrow 0$ the film thickness scales as 
$H_{\infty}/R \sim 0.643 (3 Ca)^{2/3}$ and that the plane curvature of the front and 
rear caps scales as $\kappa_{f,r} R \sim 1+\beta_{f,r}(3 Ca)^{2/3}$, with $\beta_{f,r}$ 
a different coefficient for front and rear caps. The uniform 
thin-film region is connected to the static cap of constant curvature at the 
extremities of the bubble through a dynamic meniscus \cite{cantat2013liquid}  (see Fig.~\ref{fig:geometryZones}).
The counterpart theory for a bubble in a square duct was derived 
by Wong et al.~\cite{wong1995motion1,wong1995motion2}.
However, these scalings 
agree with Taylor's experimental results 
\cite{taylor1961deposition} only in the small $Ca$ limit, namely when $Ca 
\lessapprox 10^{-3}$.
In order to understand the dynamics of confined bubbles in a 
broader parameter range, researchers have performed experiments
\cite{chen1986measuring,aussillous2000quick,fuerstman2007pressure,han2009measurement,boden2014synchrotron} as well as numerical simulations
\cite{shen1985finite,reinelt1985penetration,ratulowski1989transport,giavedoni1997axisymmetric,giavedoni1999rear,heil2001finite,kreutzer2005inertial,lac2009motion,gupta2013three,anjos20143d,anjos20143d_heat,langewisch2015prediction}. As an 
outcome, several correlations have been proposed for the evolutions of the 
relevant quantities as a function of the different parameters (see for 
example 
Ref.~\cite{han2009measurement} and Ref.~\cite{langewisch2015prediction}).
Among them,  Aussillous \& Qu\'er\'e \cite{aussillous2000quick} proposed an 
ad-hoc rational function with a fitting parameter for the film thickness which 
is in good agreement with the experimental results of Taylor \cite{taylor1961deposition}  
for capillary numbers up to $1$. The two recent works of Klaseboer et 
al.~\cite{klaseboer2014extended} and Cherukumudi et 
al.~\cite{cherukumudi2015prediction} tried to put a theoretical basis to this 
 extended Bretherton's theory for larger $Ca$.

In contrast to bubbles, which have experienced a 
vast interest of the scientific community,  little amount 
of effort has been made for droplets whose viscosities are comparable to or 
much larger than that of the outer fluid. Yet, droplets of arbitrary 
viscosities are crucial for Lab-on-a-Chip applications \cite{anna2016droplets}. 
A 
first theoretical investigation of the effect of the inner phase viscosity was conducted by Schwartz 
et al. \cite{schwartz1986motion}, motivated by the 
discrepancy in the predicted and the measured film thicknesses of long 
bubbles in capillaries. They demonstrated that the non-vanishing 
inner-to-outer viscosity ratio could thicken the film. Hodges et al. 
~\cite{hodges2004motion} further extended the theory 
and showed that the film becomes even thicker at
intermediate viscosity ratios. Numerical simulations have been performed to 
investigate the droplets in capillaries~\cite{martinez1990axisymmetric,tsai1994dynamics,lac2009motion}.

Models predicting the characteristic quantities such as the uniform film 
thickness and the meniscus curvatures of droplets in 
capillaries over a wide range of capillary numbers are still missing. For 
example, the velocity of a droplet of finite viscosity 
flowing in a channel still remains a simple question yet an open 
challenge. Such a prediction is, however, of paramount importance for the correct 
design of droplet microfluidic devices. As an example, Jakiela et 
al.~\cite{jakiela2011speed} performed extensive experiments for droplets
in square ducts, showing complex dependencies of the 
droplet velocity on the capillary number, viscosity ratio and droplet length. 
Also, what is the pressure drop induced by the presence of a drop in a channel? 
This question is crucial and has been the subjects of recent works, for example 
Refs.\ \cite{warnier2010pressure,ladosz2016pressure}. Other quantities, such as the minimum film thickness 
$H_{\textsf{min}}$, have to be accurately predicted as well. $H_{\textsf{min}}$  
becomes essential for heat transfer or cleaning of microchannels applications 
\cite{magnini2017undulations,khodaparast2017bubble}. Furthermore, being able to predict the flow field 
inside and outside of the droplet is crucial if one is interested in the mixing 
capabilities of the system.

Here, we aim at bridging this gap by combining asymptotic derivations with accurate numerical simulations
to propose physically inspired empirical models, for the 
characteristic
quantities of a droplet of arbitrary viscosity ratio flowing in an axisymmetric 
or planar capillary with a constant velocity. The model coefficients are specified by fitting laws.
The present work provides the reader with a rigorous theoretical basis, which 
can be exploited to understand the dynamics of viscous droplets. The 
considered capillary numbers vary from $10^{-4}$ to $1$ and the inner-to-outer 
viscosity ratio from $0$ to $10^5$.
Following the work of Schwartz et al.~\cite{schwartz1986motion}, we extend the 
low-capillary-number asymptotical results obtained with the lubrication 
approach of 
Bretherton \cite{bretherton1961motion} for bubbles to viscous droplets. 
Numerical simulations based on finite element method (FEM) employing the 
arbitrary Lagrangian-Eulerian (ALE) formulation are performed to validate 
the theoretical models and then extend them to the 
large-capillary-number range, $Ca\sim O(1)$, where the lubrication analysis 
fails.

The paper is structured in a way to build, step by step, 
the models for the uniform film thickness, the front and rear droplet's 
interface plane curvatures as well as those for the front and rear pressure drops 
required to compute the total 
pressure drop along a droplet in a channel. We present the problem setup, 
governing equations, numerical methods  and the validations in Sec.\ 
\ref{tx:governingEquationsNumericalMethods}. 
The flow fields inside and outside of the droplets as a function of the 
capillary numbers and viscosity 
ratios are shown in Sec.\ \ref{tx:flowField}. In particular, the flow profiles 
in the uniform-film region are derived in Sec.~\ref{tx:flowProfile} and the 
flow patterns are presented in Sec.~\ref{tx:flowPattern}. 
The theoretical part starts with the asymptotic derivation of the model for the uniform 
film thickness 
in Sec.\ \ref{tx:filmThickness}. The derivation of the lubrication equation is 
detailed in Sec.~\ref{tx:filmThicknessSmallCaImplicit}, followed by the film 
thickness model in Sec.~\ref{tx:filmThicknessSmallCaExplicit} and its extension 
to larger capillary numbers in Sec.~\ref{tx:filmThicknessLargerCa}.
With the knowledge of the film thickness, the droplet velocity can be computed 
analytically (see Sec.\ \ref{tx:dropletVelocityUinf}). The minimum film 
thickness separating the droplet form the channel walls is discussed in Sec.\ 
\ref{tx:minFilmThickness}. To build a total pressure drop model, one still needs 
the knowledge of the front and rear caps mean curvatures (see Sec.\ 
\ref{tx:frontRearCurvatures}), the front and rear pressure jumps (see Sec.\ 
\ref{tx:frontRearPJumps} and \ref{tx:frontRearPJumpsImproved}) and the front and 
rear normal viscous stress jumps (see Sec.\ \ref{tx:frontRearUzJumps}). The 
stresses evolutions at the channel centerline and at the wall are presented in 
Sec.\ \ref{tx:stressesEvolCenterline} and Sec.\ \ref{tx:stressesEvolWall}, 
respectively. Eventually, one can sum up all these contributions to build the 
total pressure drop, which is described in Sec.\ \ref{tx:totalPressureDrop}. 
We summarize our results in Sec.\ \ref{tx:conclusions}.

\section{Governing equations and numerical methods \label{tx:governingEquationsNumericalMethods}}

\subsection{Problem setup \label{tx:consideredProblem}}

\begin{figure}[ht!]
\includegraphics[width=0.45\textwidth]{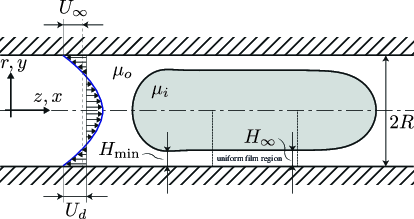}
\caption{Sketch of the axisymmetric ($z,r$) and planar ($x,y$) 
configurations in the frame of reference moving with the droplet. The flow 
profiles in the uniform film region are shown in Fig.\ \ref{fig:velocityProfiles}. 
\label{fig:geometry}}
\end{figure}

We consider an immiscible droplet of volume 
$\Omega$ and dynamic viscosity $\mu_i$ translating at a steady velocity $U_d$ 
in a planar channel/axisymmetric tube of width/diameter $2R$
filled with an outer fluid of dynamic viscosity $\mu_o$.
The volume flux of the outer fluid is $q_o$, resulting in  an average 
velocity $U_{\infty}=q_o/(2R)$ and
$U_{\infty}=q_o/(\pi R^2)$ for the planar and 
axisymmetric configuration, respectively (see Fig.~\ref{fig:geometry}). Given 
the small droplet velocity and size, the Reynolds number $Re$ is small and 
inertial effects can be neglected. Buoyancy is also neglected. The relevant
dimensionless numbers include the droplet capillary number $Ca = \mu_o 
U_d/\gamma$ with $\gamma$ being the surface tension of the droplet interface 
and the dynamic viscosity ratio $\lambda = 
\mu_i/\mu_o$ between the droplet and the outer fluid. The capillary number based on 
the mean flow velocity is $Ca_{\infty}=\mu_o U_{\infty}/\gamma$.
For the numerical simulations, the droplet capillary number has been varied within 
$10^{-4}\lessapprox Ca \lessapprox 1$ to guarantee that
the lubrication film is only influenced by the hydrodynamic forces and
the dynamics is steady.
For smaller capillary numbers, non-hydrodynamic forces such as disjoining 
pressures due to intermolecular forces might come into play as reported by the 
recent experiments~\cite{huerre2015droplets}; while for larger capillary 
numbers, instability and unsteadiness might arise, where the droplet might 
form a re-entrant 
cavity at its rear \cite{tsai1994dynamics}, which eventually breaks up 
into satellite droplets 
The viscosity ratios investigated numerically are from the 
well-known bubble limit of $\lambda=0$~\cite{bretherton1961motion} to 
highly-viscous droplets of $\lambda=100$ that has been scarcely investigated.

We consider both a three-dimensional axisymmetric tube, and a 
two-dimensional planar
channel featured by the span-wise invariance. Note 
that the latter configuration does not correspond to the Hele-Shaw-cell-like 
microfluidic chips, where Darcy or Brinkman equations are more appropriate to
describe the flow
\cite{boos1997thermocapillary,nagel2015boundary}.

It is worth noting that the confined droplet has to be long enough to 
develop a region of uniformly thick film at its center 
\cite{cantat2013liquid} (see Fig.~\ref{fig:geometry}). However, a long axisymmetric droplet is likely to become unstable to the Rayleigh-Plateau instability. The uniform film region would resemble to a coaxial jet, which is known to be unstable to perturbations with a wavelength longer than $2\pi(R-H_{\infty})$ \cite{eggers2008physics}.
Within this range of droplet lengths, we 
found that the effect of droplet volume $\Omega$ is insignificant 
and hence it is fixed to $\Omega/R^3=12.9$ for the axisymmetric geometry 
and $\Omega/R^2=9.3$ for the planar case.

\subsection{Governing equations \label{tx:governingEquations}}

The governing equations are the incompressible Stokes equations for the velocity $\vec{u}=(u,v)$ and pressure $p$: 
\begin{align}
	\nabla \cdot \vec{u} = & 0 \label{eq:StokesContinuity}\\
	\vec{0} = & \nabla \cdot \boldsymbol{\sigma},
	\label{eq:StokesMomentum}
\end{align}
where $\boldsymbol{\sigma} = - p \tens{I} + \mu \left[ \left( \nabla 
\vec{u} \right) + \left( \nabla \vec{u} \right)^T \right]$ denotes the 
total stress tensor and $\mu$ the dynamic viscosity as $\mu_i$ (resp. 
$\mu_o$) inside (resp. outside) the droplet.

The imposed dynamic boundary conditions at the interface are the continuity of 
tangential stresses 
\begin{equation}\label{eq:bcTang}
	\Delta \left[(\tens{I}-\bn\bn) \cdot \left( \tens{\sigma} \cdot 
\vec{n}   \right)\right] = \vec{0} ,
\end{equation}
and the discontinuity of normal stresses due to the Laplace pressure jump
\begin{equation}\label{eq:bcNorm}
	\Delta \left( \tens{\sigma} \cdot \vec{n}  
 \right) = - \gamma \mathcal{C} \bn.
\end{equation}
 $\Delta$ denotes the difference between the inner and outer quantities, 
$\vec{n}$ the unit normal vector on the interface towards the outer fluid, 
and $\mathcal{C} = \nabla_S \cdot \bn $ the interfacial mean curvature ($\nabla_S$ is 
the surface gradient). The plane curvature of interface in the ($z,r$) or ($x,y$) plane is denoted as $\kappa$ and its value at the front and rear droplet extremities is given by $\kappa_f$ and 
$\kappa_r$, respectively. The mean curvature at the front and rear droplet extremities, which lie on the symmetry axis, is therefore given by $\mathcal{C}_{f,r} = \chi \kappa_{f,r}$, where $\chi = 1$ or $2$ for a planar or axisymmetric configuration, respectively. At any other point on the droplet interface, the mean curvature is given by the sum of the two principal curvatures.

\subsection{Numerical methods \label{tx:numericalMethodsSolver} and 
implementations}

\begin{figure}[ht!]
\includegraphics[width=0.45\textwidth]{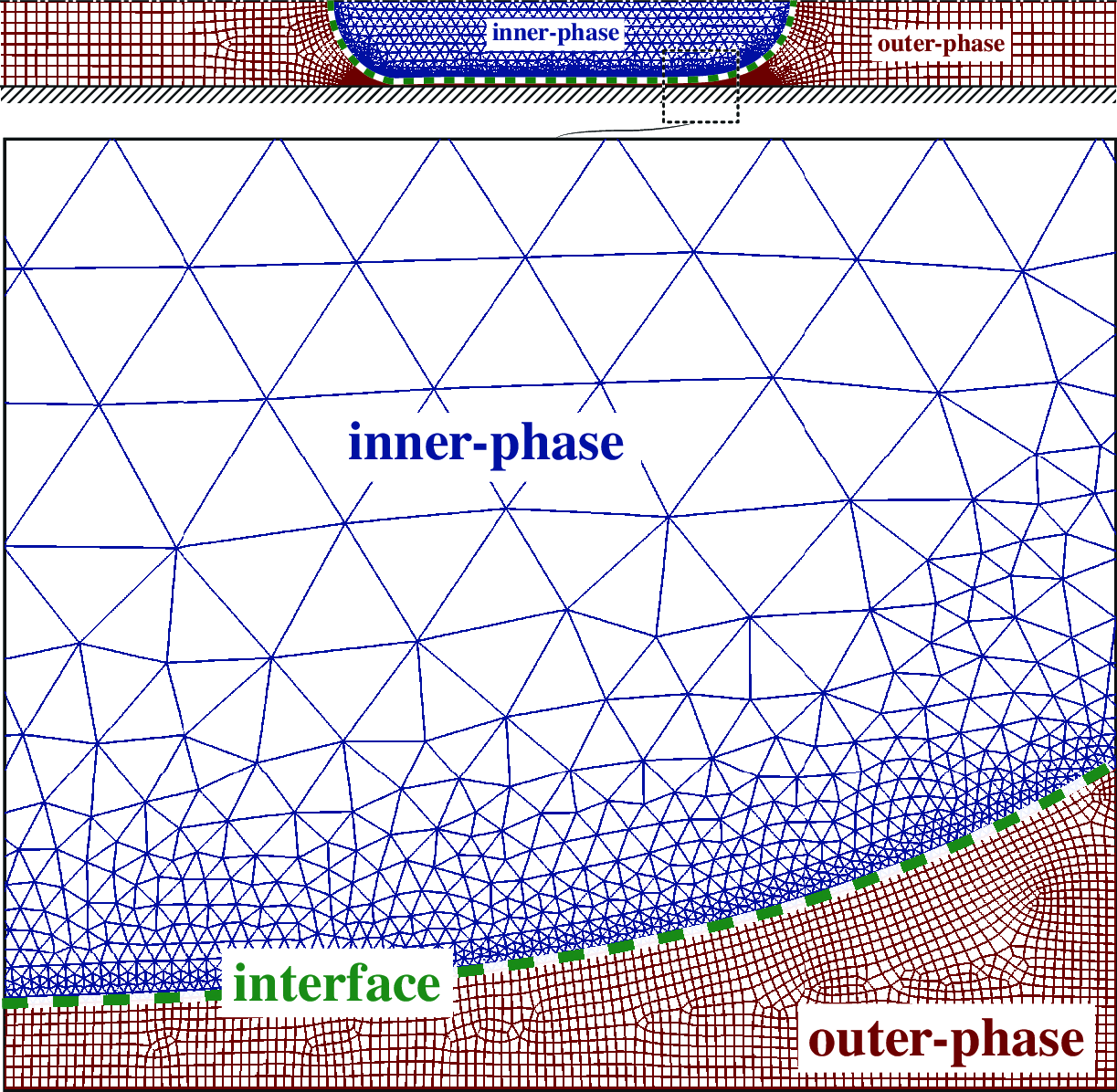}
\caption{Computational mesh. Inset: mesh refinement in the thin-film 
region. The triangular inner-phase (blue) and quadrilateral outer-phase meshes 
(red) 
are separated by the explicitly discretized interface (dashed green).\label{fig:mesh}}
\end{figure}

Equations \eqref{eq:StokesContinuity}-\eqref{eq:StokesMomentum} 
with boundary conditions \eqref{eq:bcTang}-\eqref{eq:bcNorm} are solved by the commercial FEM package
COMSOL Multiphysics and the interface is well represented by the 
arbitrary Lagrangian-Eulerian (ALE) technique. Compared to the 
commonly known diffuse interface methods such as volume-of-fluid, phase-field, 
level-set and front-tracking all relying on a fixed Eulerian grid, the ALE 
approach captures the interface more accurately. Since the 
interface is always explicitly represented by the discretization points (see 
Fig.~\ref{fig:mesh}), the fluid quantities (viscosity, density, 
etc.), pressure and normal viscous stresses, are discontinuous across the 
interface.
This technique has been used to simulate three-dimensional bubbles in complex 
microchannels~\cite{anjos20143d}, liquid films coating the interior of 
cylinders \cite{hazel_heil_waters_oliver_2012},
two-phase flows with surfactants~\cite{ganesan2017ale,ganesan2012arbitrary} and
head-on binary droplet collisions~\cite{li2016macroscopic}, to name a few. The \textit{Moving Mesh} module of COMSOL Multiphysics has been recently employed in Refs.~\cite{rivero2018bubble,hadikhani2018inertial} to investigate the inertial and capillary migration of bubbles in round and rectangular microchannels.

Despite the high fidelity in interface capturing, it is commonly more challenging 
to develop in-house ALE implementations compared to the diffuse interface 
counterparts. Additional difficulty arises in the case of large interfacial 
deformations,
when the computational mesh might become highly nonuniform and skewed. It is therefore necessary to remesh the computational domain and to obtain all the quantities on the new mesh via interpolation.
Hence, special expertise in scientific computing and 
tremendous amount of development effort is required to implement an in-house 
ALE-based multi-phase flow solvers, which have prevented large 
portion of the research community from enjoying the high fidelity and elegance 
of the ALE methods. 

Hereby, large mesh deformations can be avoided not only thanks to the convenient moving mesh module of COMSOL, but also by
solving the problem in the moving frame of droplet. To achieve so, we impose
a laminar Poiseuille inflow of mean velocity $U_{\infty}-U_d$
at the inlet of the 
channel and the velocity $-U_d$ at the walls, where the unknown droplet velocity 
$U_d$ is obtained as part of the solution together with that of the flow field, 
at each time step, 
by applying an extra constraint of zero
volume-integrated velocity inside the droplet. Such constraints
with additional unknowns are imposed in COMSOL Multiphysics by utilizing its
so-called 'Global 
Equations'. This strategy 
ensures that the deforming droplet barely translates in the streamwise 
direction, staying approximately at its initial position (say in the center 
of the domain). Hence, the mesh quality and the robustness of the ALE 
formulation is appropriately guaranteed.

In this work, we are only concerned with the steady dynamics of the droplet
reaching its equilibrium shape.
We do not solve the steady Stokes 
Eq.~\eqref{eq:StokesMomentum} strictly but maintain a negligible time-derivative term 
$Re \frac{\partial \vec{u}}{\partial 
t}$ for 
time marching. This procedure can be seen as an iterative scheme to find steady solutions of the Stokes equations.
Since, indeed, the time-derivative term vanishes when the equilibrium state is reached, the solutions to the steady Stokes equations are eventually obtained.

It has to be stressed the computed transient dynamics is not physical and only the final steady solution should be considered. Theoretically, the stability of the latter might be affected by the time-derivative term. 
However, in practice, no unstable 
phenomena or multiple-branch solutions were observed
in our study.
We monitored, during each simulation, the temporal 
evolution of the velocity, the mean and minimum film thickness as well as the front and rear 
plane curvatures of the droplet, which exhibited precise 
time-invariance without exception when the equilibrium state was reached.
Our method therefore converges to valid stationary solutions, which will be seen in Sec.~\ref{tx:numericalMethodsValidation} to compare well with asymptotic estimates and numerical solutions from the literature.

To reduce the computational cost, half of the channel is considered and 
axisymmetric or symmetric boundary conditions are imposed at the channel 
centerline for the axisymmetric and planar configurations, respectively.
The setup in COMSOL Multiphysics is intrinsically parallel and the computing time required 
for an individual case needs no more than one hour based on a 
standard desktop computer.

A typical mesh is shown in Fig.\ \ref{fig:mesh}. Triangular/quadrilateral 
elements are used to discretize the domain inside/outside the droplet. 
Furthermore, a mesh refinement is performed to best resolve the thin 
lubrication film (see inset of Fig.~\ref{fig:mesh}). It is worth-noting that 
quadrilateral elements have to be used to discretize
the thin film because this region might undergo large radial deformation 
resulting
in highly distorted and skewed triangular elements if used.

\subsection{Validation \label{tx:numericalMethodsValidation}}

\begin{figure}[ht!]
\includegraphics[width=0.45\textwidth]{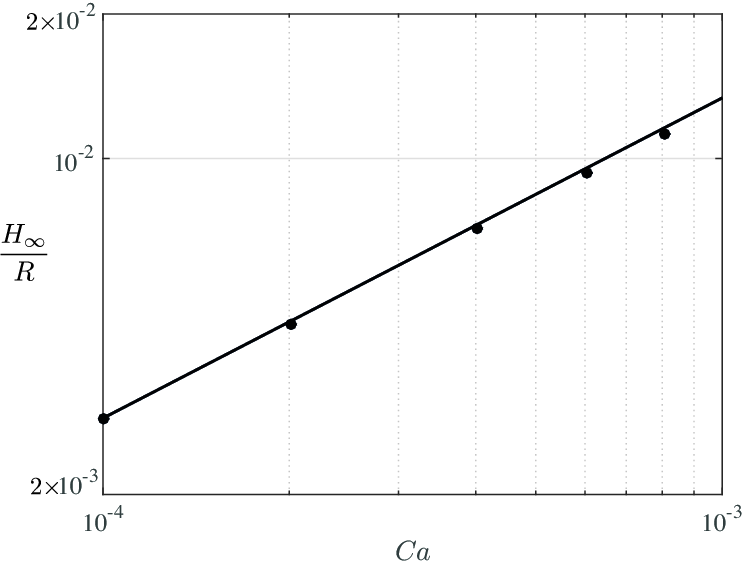}
\caption{Comparison between the uniform film thickness between the wall and a bubble obtained by the
FEM-ALE simulations 
(symbols) 
and that predicted by Bretherton \cite{bretherton1961motion} (solid line) for 
the planar 
channel.\label{fig:validationBretherton}}
\end{figure}

\begin{figure}[ht!]
\includegraphics[width=0.45\textwidth]{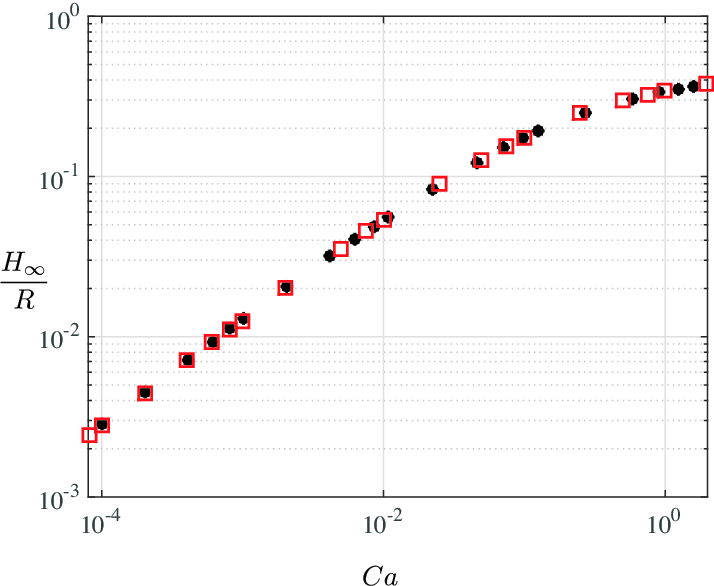}
\caption{Comparison between the uniform film thickness between the wall and a bubble obtained by 
the FEM-ALE simulations 
(full black circles) 
and that of Ref.\ \cite{giavedoni1997axisymmetric} (empty red squares) for the planar 
channel.\label{fig:validation}}
\end{figure}

Our numerical results are first validated for a bubble ($\lambda=0$) comparing 
the film thickness with the classical asymptotic theory $H_{\infty}/R \sim 0.643 (3 Ca)^{2/3}$ of Bretherton in the low-$Ca$ limit 
\cite{bretherton1961motion} (see 
Fig.~\ref{fig:validationBretherton}). Excellent agreement is revealed even when 
the capillary number is $10^{-4}$; the discrepancy at larger 
$Ca$ is mostly because of the asymptotic nature of the model that becomes less
accurate for increasing $Ca$. At 
larger capillary numbers, we compare the uniform film thickness with the 
FEM-based numerical results of 
Ref.~\cite{giavedoni1997axisymmetric} for a bubble, showing perfect agreement
in Fig.\ \ref{fig:validation}; agreements for the front and rear 
plane curvatures are also observed and are not reported here. 

\begin{figure}[ht!]
\includegraphics[width=0.45\textwidth]{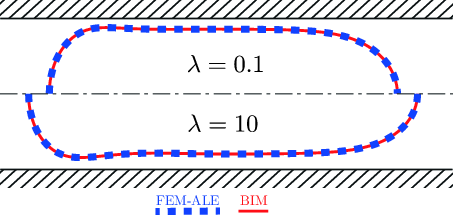}
\caption{Comparison between the droplet profiles obtained by the FEM-ALE
computations (blue dashed)  and the BIM  (red solid) computations of 
Ref.~\cite{lac2009motion} for an axisymmetric droplet in a tube with 
$Ca_{\infty}=0.05$ of viscosity ratios $\lambda=0.1$ (upper half domain) and 
$\lambda=10$ (lower half domain).}
\label{fig:validation_lac}
\end{figure}

For  viscosity ratios $\lambda >0$, we have validated our 
setup against the results from an axisymmetric boundary integral method (BIM)
solver~\cite{lac2009motion} for a droplet with 
$Ca_{\infty}=0.05$  of viscosity ratios $\lambda=0.1$ 
and $10$, again exhibiting perfect agreement as displayed in 
Fig.~\ref{fig:validation_lac}.

Based on the carefully performed validations against the theory,
numerical results from FEM and BIM solvers, we are confident that
the developed COMSOL implementation can be used to carry out high-fidelity 
two-phase simulations efficiently, at least for the $2$D and $3$D-axisymmetric
configurations.


\section{Flow field \label{tx:flowField}}

\subsection{Velocity profiles in the thin-film region 
\label{tx:flowProfile}}
\begin{figure}[ht!]
\includegraphics[width=0.45\textwidth]{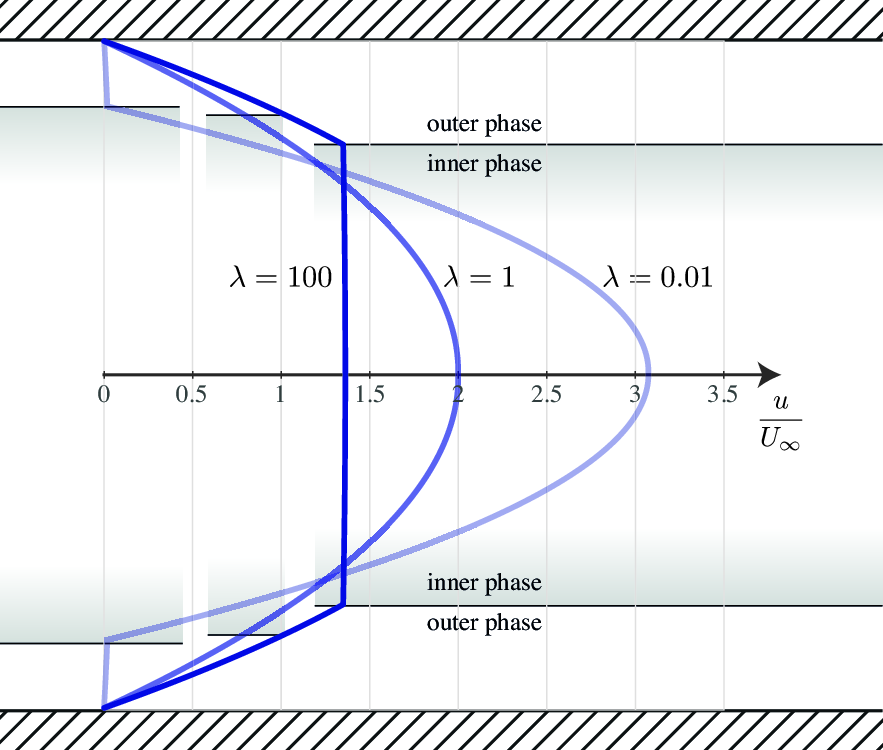}
\caption{Inner and outer phase velocity profiles in the uniform film region of 
an axisymmetric droplet with $Ca_{\infty} = 
0.1$ and viscosity ratios $\lambda = 0.01$, $1$ and $100$, represented in the  
laboratory frame.
 \label{fig:velocityProfiles}}
\end{figure}

For a sufficiently long droplet/bubble, 
a certain portion of the lubrication 
film is of uniform thickness $H_{\infty}$ \cite{bretherton1961motion}
(see Fig.\ \ref{fig:geometry} and Fig.~\ref{fig:streamlines}). Within this portion, 
the velocity field both inside and outside the droplet is invariant in the 
streamwise direction and resembles the well known bi-Poiseuille profile that 
typically arises in several interfacial flows,
for example a coaxial jet~\cite{herrada2008spatiotemporal} (see Fig.\ 
\ref{fig:velocityProfiles}). For $\lambda \ll 1$, the velocity profile in the 
film is almost linear, whereas for $\lambda \gg 1$, the velocity 
inside of the droplet is almost constant (plug-like profile). 
Nevertheless, the 
parabolic component of these profiles is crucial for the accurate prediction of 
the droplet velocity (see Sec.\ \ref{tx:dropletVelocityUinf}).

Assuming the bi-Poiseuille velocity profile, we describe the streamwise 
velocity $u_i(r)$ inside and $u_o(r)$ outside the droplet as a function of the 
off-centerline distance $r$ as:
\begin{align}
	u_i(r) &= \frac{1}{4 \mu_i} \frac{d p_i}{dz} r^2 + A_i \ln r + B_i, \\
	u_o(r) &= \frac{1}{4 \mu_o} \frac{d p_o}{dz} r^2 + A_o \ln r + B_o,
\end{align}
where $p_i$ and $p_o$ are the inner, respectively outer, pressures, 
and $A_i$, $B_i$, $A_o$ and $B_o$ are undetermined constants. 
Given the finiteness of $u_i(r)$ at $r=0$, we have $A_i=0$. By satisfying the 
no-slip boundary condition on the 
channel walls $u_o(R) = -U_d$, the continuity of 
velocities and tangential stresses on the interface $r=\tilde{r}=R-H$, 
namely, $u_i(\tr) = u_o(\tr)$ and 
\begin{equation}
	\mu_i \frac{d u_i}{dz}\bigg\vert_{r=\tr} = \mu_o \frac{d 
u_o}{dz}\bigg\vert_{r=\tr},
\end{equation}
we obtain the remaining constants
\begin{align}
	A_o =& \frac{1}{2 \mu_o} \left(\frac{d p_i}{dz}-\frac{d p_o}{dz}\right)
	\tr^2, \\
	B_i =& -\frac{1}{4 \mu_i \mu_o} \left[ \frac{d p_o}{dz} (R^2 - \tr^2)  
\mu_i +  \frac{d p_i}{dz}  \tr^2  \mu_o \right. \\
	       &+ \left. 2 \left(\frac{d p_i}{dz} - \frac{d p_o}{dz}\right) 
\tr ^2 \mu_i \ln \left( \frac{R}{\tr}\right) \right] -U_d, \nonumber \\
	B_o =& -\frac{1}{4 \mu_o}\left[ \frac{d p_o}{dz} R^2 +2\left(\frac{d 
p_i}{dz}-
\frac{d p_o}{dz}\right)\tr^2 \ln R \right] -U_d.
\end{align}
Under the assumption of a slowly evolving film thickness, this velocity profile also holds in the nearby regions, where the thickness is $H$ rather than $H_{\infty}$. The derivation for the planar geometry is given in Appendix \ref{tx:flowProfilesPlanar}.

\subsection{Recirculating flow patterns \label{tx:flowPattern}}

The axisymmetric velocity profile in the channel away from the droplet, in its moving reference frame, is given by
\begin{equation}
	u_{\infty}(r) = 2 U_{\infty}\left[1-\left(\frac{r}{R}\right)^2\right]-U_d.
	\label{eq:uInf}
\end{equation}
As will become clear in Sec.~\ref{tx:dropletVelocityUinf}, the droplet velocity 
can be obtained by imposing mass conservation. For the particular case of a 
bubble with $\lambda = 0$, mass conservation reads
$(U_{\infty}-U_d)\pi R^2 = - \pi[R^2-(R-H_{\infty})^2] U_d$ 
\cite{stone2010interfaces}, yielding $U_{\infty}/U_d = 
(1-H_{\infty}/R)^2$. The velocity profile \eqref{eq:uInf} can therefore be  
expressed as a function of $H_{\infty}/R$. As pointed out by Giavedoni \& Saita 
\cite{giavedoni1997axisymmetric}, when $\lambda=0$, the velocity at the 
centerline $u_{\infty}(r=0)$ in the bubble frame  
changes sign when $H_{\infty}=\Hc=(1-1/\sqrt{2})R$. As a consequence, when 
the uniform film thicknesses is below $\Hc$, $u_{\infty}(0)>0$, an external 
recirculating flow pattern forms ahead of and behind the bubble. For the planar configuration, 
the critical film thickness for the appearance of the flow recirculation is 
$\Hc = R/3$.

Based on the flow profiles derived in Sec.~\ref{tx:flowField}
and mass conservation (see Sec.~\ref{tx:dropletVelocityUinf}), we can generalize the critical thickness $\Hc$ to 
non-vanishing viscosity ratios ($\lambda>0$) as (see Appendix \ref{tx:derivationHstar} for the derivations):
\begin{equation}
	\frac{H_{\infty}^{\star}}{R} = 1-\sqrt{\frac{(\lambda -1)(2 \lambda -1)}{2}}\frac{1}{\lambda -1}
	\label{eq:HcAxi}
\end{equation}
for the axisymmetric case and
\begin{equation}
	\frac{H_{\infty}^{\star}}{R} = \frac{1}{3}\frac{1}{1-\lambda}
	\label{eq:HcPlanar}
\end{equation}
for the planar case.  
$\Hc/R$ reaches the value of $1$ when $\lambda = 1/2$ or $\lambda = 2/3$, for the axisymmetric and planar configuration, 
respectively. Nevertheless,
the uniform film thickness is always much smaller than the channel half-width, $H_{\infty}\ll R$.
For a fixed volume of fluid, large film thicknesses would correspond to long droplets, which might be unstable to the Rayleigh-Plateau instability as discussed in Sec.\ \ref{tx:consideredProblem}.

For viscosity ratios $\lambda \geq 1/2$ ($\lambda \geq 2/3$) for
the axisymmetric (planar) configuration, there is no critical film thickness 
above which the recirculation zones disappear, meaning that a 
recirculation region always exists for any capillary number. 

\begin{figure}
\includegraphics[width=0.48\textwidth]{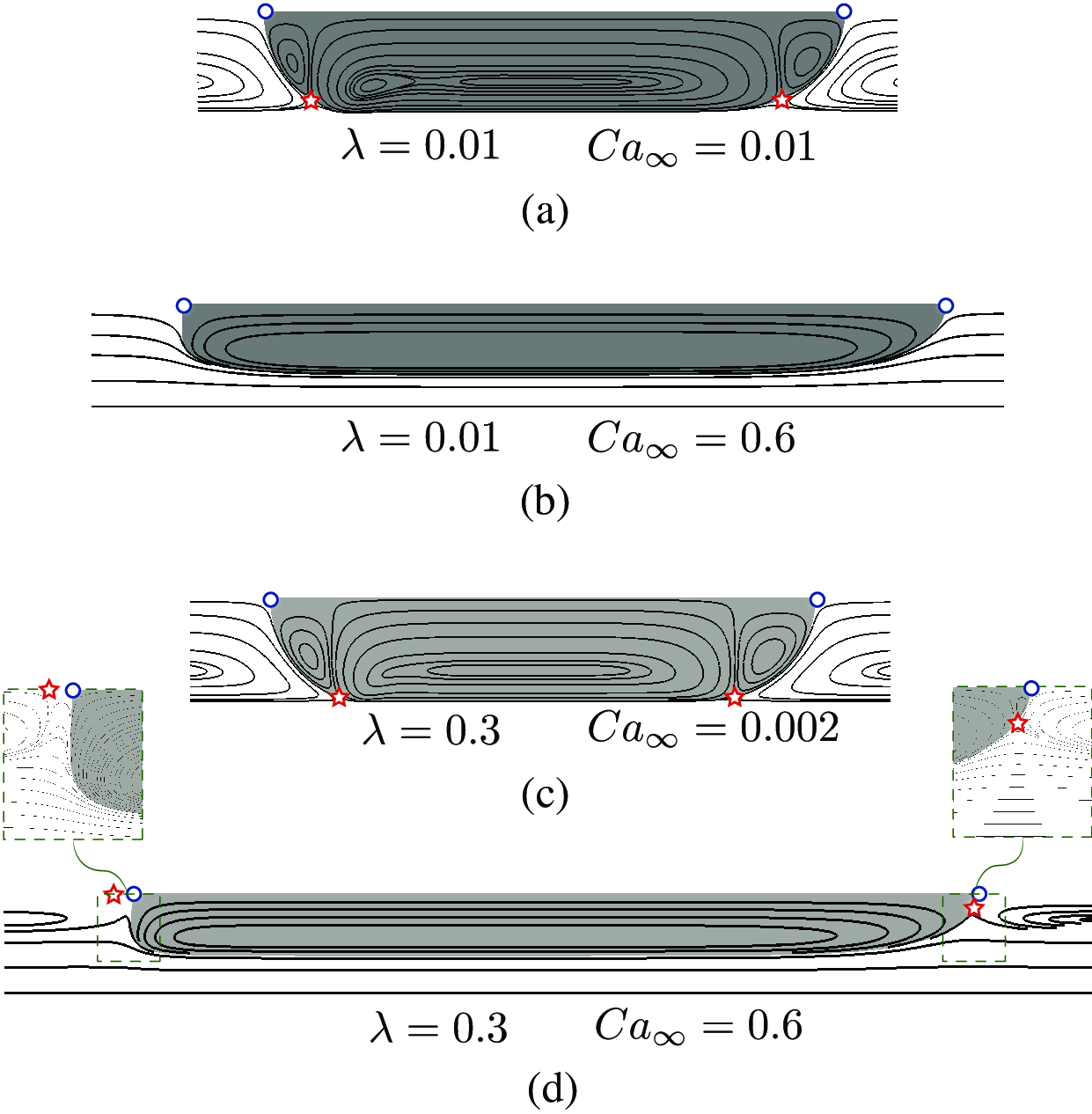}
\caption{Streamlines and recirculation patterns for an axisymmetric droplet with
different capillary numbers 
$Ca_{\infty}$ and viscosity ratios $\lambda$ in a 
frame of reference moving with the droplet. The results are obtained from 
FEM-ALE numerical simulations. The permanent stagnation points at the droplet vertices are 
highlighted by blue circles, whereas the ($Ca$,$\lambda$)-dependent stagnation rings/points are highlighted by red stars. Insets: detailed flow pattern at the droplet front and rear for 
$\lambda = 0.3$ and $Ca_{\infty}=0.6$.
\label{fig:streamlines}}
\end{figure}

At low capillary numbers, when the film thickness is below 
$H_{\infty}^{\star}$, the 
external recirculating flows are strong enough to induce 
the recirculation inside the droplet as well. Consequently, besides the 
two droplet vertices as permanent stagnation points (blue circles in Fig.~\ref{fig:streamlines}), two 
stagnation rings emerge on the front and rear part of the 
axisymmetric interface (red stars in Fig.~\ref{fig:streamlines}); 
likewise, four stagnation points arise in the planar 
case. 
The stagnation rings/points on the dynamic meniscus (red stars in 
Fig.~\ref{fig:streamlines}) move outwards to the droplet vertices 
when $Ca$ increases. When $H_{\infty}>\Hc$, these 
stagnation rings/points disappear, taking away with the recirculation regions 
accordingly (see Fig.\ \ref{fig:streamlines}(b)). 
Only the stagnation points at the droplet vertices remain.

However, since the stagnation rings/points at the droplet interface move 
outwards to the front and rear extremities when the film thickness increases 
with the capillary number,
the recirculation regions might eventually detach from the interface before the 
critical film thickness ${H_{\infty}^{\star}}$ is reached. Nevertheless, 
since recirculation regions exist far away from the droplet as long as 
$H_{\infty}<\Hc$, another flow pattern must exist close to the droplet 
interface. The detached stagnation 
points are located on the centerline outside the droplet, 
and no recirculating flow is present in the region between the detached 
stagnation point and the one at the droplet vertex (see rear of droplet in 
Fig.~\ref{fig:streamlines}(d)). 
Contrasting with the case of Fig.~\ref{fig:streamlines}(c), 
where the stagnation points on the dynamic meniscus
induce recirculating flow inside the droplet at 
both front and rear parts,
the inner rear recirculation zone disappears when the capillary number increases, as 
shown in  Fig.~\ref{fig:streamlines}(d). This indicates that
the detachment of the 
recirculation region is not fore-aft symmetric.
There is a large range of 
parameters for which a rear stagnation point is not at the droplet interface 
anymore and thus there is no recirculation inside the rear part of the 
droplet. We have found that the critical film thickness for which the stagnation 
ring/point at the rear detaches from the droplet interface corresponds to 
when the rear plane curvature of the droplet changes the sign (see also 
Sec.~\ref{tx:frontRearCurvatures}). For both the flow patterns as well as for 
the plane curvatures, the fore-aft asymmetry increases with 
the capillary number.

\begin{figure}
\includegraphics[width=0.45\textwidth]{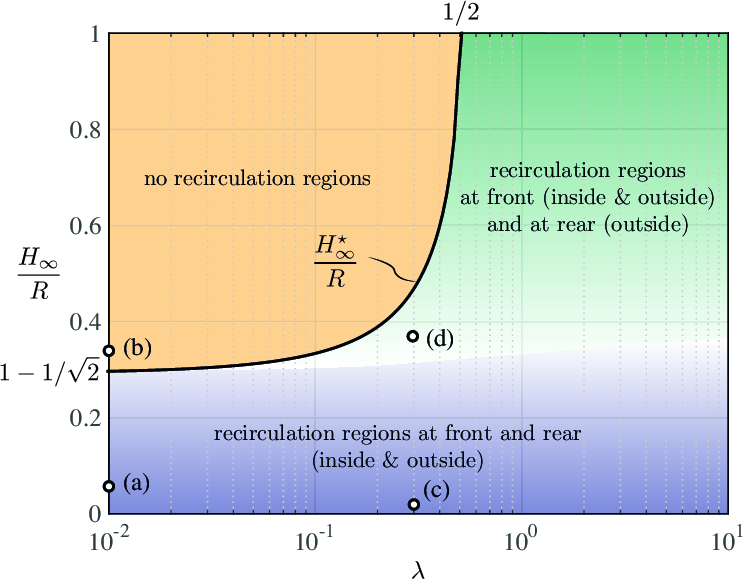}
\caption{Diagram of the main possible flow patterns for the axisymmetric 
configuration. 
The streamlines corresponding to the points (a)-(d) are shown in 
Fig.\ 
\ref{fig:streamlines}.\label{fig:plotStagnationPointDiag}}
\end{figure}

The phase diagram with the main different types of flow patterns as a function of the viscosity
ratio $\lambda$ and film thickness $H_{\infty}/R$ is shown in Fig.\ 
\ref{fig:plotStagnationPointDiag}. Note that for viscosity ratios $\lambda \geq 
1/2$ 
($\lambda \geq 2/3$), the recirculation regions will be attached or 
detached from or to the droplet interface depending on the the uniform 
film thickness.
Other very peculiar flow fields, as a 
detached finite recirculation region at the rear or a detached recirculation 
region at the front as observed by Giavedoni \& Saita 
\cite{giavedoni1997axisymmetric,giavedoni1999rear}, can be obtained for some 
parameter combinations. However, since the flow field structures are not the 
main aim of this work, an extended parametric study to detect all possible 
patterns has not been performed. Nevertheless, our current results 
do not validate all the flow patterns predicted in 
Ref.~\cite{hodges2004motion} based on
asymptotic arguments, which indeed have not been verified neither experimentally nor 
numerically.                               

The flow fields for the planar configuration are not presented here as they are
similar to those for the axisymmetric configuration.

\section{Film thickness \label{tx:filmThickness}}

\subsection{Asymptotic result in the \textit{low}-$Ca$ limit 
\label{tx:filmThicknessSmallCaImplicit}}
By following the work of Schwartz et al.~\cite{schwartz1986motion},
we derive an implicit expression predicting the thickness
$H_{\infty}$ of the uniformly-thick lubrication film in the low-$Ca$
limit when $H/R \ll 1$ satisfies. The derivation of the axisymmetric
 case is presented below, see Appendix \ref{tx:derivationSchwartzPlanar} for the 
planar case.

The flow rates at any axial location where the external film thickness is $H$ are:
\begin{align}
	q_i =& 2 \pi \int_0^{R-H} u_i(r) r dr \label{eq:QiFull} \\
	   = & -\pi ( R-H)^2 \left\{ U_d + \frac{1}{8 \mu_i \mu_o} \left[2 \frac{d p_o}{dz} H (2 R - H) \mu_i  \right. \right.  \nonumber \\
	    + & \frac{d p_i}{dz}(R-H)^2\mu_o\nonumber  \nonumber \\
	   +& \left. \left. 4 \left(\frac{d p_i}{dz}-\frac{d p_o}{dz}\right) (R-H)^2 \mu_i \ln \left(\frac{R}{R-H}\right)\right] \right\}, \nonumber \\
	q_o =& 2 \pi \int_{R-H}^{R} u_o(r) r dr \label{eq:QoFull}\\
	        =&- \frac{\pi}{8 \mu_o}\left\{ H(2R-H) \left[H^2\left(2\frac{d p_i}{dz} -3\frac{d p_o}{dz}\right) \right. \right. \nonumber \\
	        +& \left. 2\left( \frac{d p_i}{dz}-\frac{d p_o}{dz}\right) R^2 -H\left(4\frac{d p_i}{dz} - 6\frac{d p_o}{dz}\right)R\right]   \nonumber \\ 
	        + &  \left. 4 \left(\frac{d p_i}{dz}-\frac{d p_o}{dz}\right) (R-H)^4 \ln \left(\frac{R}{R-H}\right)\right\}  \nonumber \\
	        -& \pi H(2R-H) U_d . \nonumber
\end{align}
Assuming that  $H/R \ll 1$,
the volumetric fluxes up to the second order are
\begin{align}
	q_i \approx &- \pi R^2 \left( U_d + \frac{1}{8 \mu_i}\frac{d p_i}{dz} R^2 + \frac{1}{2 \mu_o}\frac{d p_i}{dz} R H + \frac{1}{2 \mu_o}\frac{d p_o}{dz} H^2 \right), \label{eq:Qi} \\
	q_o \approx &- 2 \pi R H \left( U_d + \frac{1}{4 \mu_o}\frac{d p_i}{dz}H R + \frac{1}{3 \mu_o}\frac{d p_o}{dz} H^2 \right). \label{eq:Qo}
\end{align}
In the droplet frame, the inner flow rate is $q_i =0$. 
Furthermore, in the region with a uniformly-thick film, 
$H=H_{\infty}$; the inner and outer pressure gradient balances, $\frac{d 
p_i}{d z}=\frac{d p_o}{d z}$. Using 
these two conditions one can obtain the pressure gradient in the uniform film 
region
\begin{equation}
	\frac{d p}{dz} \bigg\vert_{r=R-H_{\infty}} \approx -\frac{8 \mu_i  U_d}{R^2+4 \lambda H_{\infty} R + 4 \lambda H_{\infty}^2}
\end{equation}
and the outer flow rate in the $H_{\infty}/R \ll1 $ limit is
\begin{align}
	q_o \approx & -2 \pi R H_{\infty} \left[\frac{3 R^2  + 6 \lambda H_{\infty}  R +4 \lambda H_{\infty}^2 }{
 3 (R^2  + 4 \lambda H_{\infty}  R+4 \lambda H_{\infty}^2)}\right]U_d \nonumber \\ 
   \approx &- 2 \pi R H_{\infty} \left(\frac{ R + 2 \lambda H_{\infty}   }{ R  + 4 \lambda H_{\infty}  }\right)U_d.
   \label{eq:qoThinFilm}
\end{align}
In the dynamic meniscus regions, the inner and outer pressure gradients are not 
equal and their difference is proportional to the mean curvature of the 
interface at $r=R-H$, through Laplace's law. 
By assuming a quasi-parallel flow whose radial velocity is weaker
than the axial velocity by one order of magnitude and a small capillary
number,
the normal stress condition at the interface is not affected by viscous 
stresses. Hence, the axial gradient of the normal stress condition reads
\begin{equation}
	\frac{d p_i}{dz} - \frac{d p_o}{dz} = \gamma \frac{d^3 H}{d 
z^3},
	\label{eq:Laplace}
\end{equation}
where the axial gradient of the curvature in the azimuthal direction is 
neglected as it is an order smaller.
The pressure gradients $dp_i/dz$ and $dp_o/dz$ as functions 
of $H$ can be obtained from  Eqs.\ \eqref{eq:Qi}, \eqref{eq:Qo} imposing mass conservation, \textit{i.e.} imposing $q_i=0$ and $q_o$ given by Eq.~\eqref{eq:qoThinFilm}:
\begin{align}\label{eq:dpi_dz}
 \frac{d p_i}{dz} \approx & \frac{4 \lambda (-6 H_{\infty}^2 \lambda + 4 
H_{\infty} H \lambda - 3 H_{\infty} R + H R) \mu_o U_d }{H R (4 H_{\infty} 
\lambda + R) (H \lambda + R) },  \\
\label{eq:dpo_dz}  \frac{d p_o}{dz} \approx & \frac{3 (H_{\infty} - H) [8 H_{\infty} H \lambda^2 
+ 2 (H_{\infty} + H) \lambda R + R^2] 
\mu_o U_d}{H^3 (4 H_{\infty} \lambda + R) (H \lambda + R)}.
\end{align}
By plugging Eqs.~\eqref{eq:dpi_dz}, \eqref{eq:dpo_dz} into Eq.\ \eqref{eq:Laplace} and adopting the 
change of variables $H = H_{\infty} \eta$ and $z = H_{\infty} (3 
Ca)^{-1/3} \xi$ in the spirit of Bretherton 
\cite{bretherton1961motion},
we obtain an universal governing equation for the scaled film thickness $\eta$ 
when taking the limit $H_{\infty}/R\rightarrow 0$:
 \begin{equation}
 	\frac{d^3 \eta}{d\xi^3} = \frac{\eta -1}{\eta^3}\left[\frac{1+2 m (1+ \eta + 4 m \eta)}{(1+4m)(1+m\eta)}\right],
	\label{eq:SchwartzAxi}
 \end{equation}
 where 
\begin{equation}
	m = \lambda \frac{H_{\infty}}{R}\nonumber
	\label{eq:m}
\end{equation}
denotes the rescaled viscosity ratio. The corresponding planar counterpart 
reads (see derivation in Appendix \ref{tx:derivationSchwartzPlanar})
  \begin{equation}
 	\frac{d^3 \eta}{d\xi^3} =2\frac{\eta -1}{\eta^3}\left[ \frac{2+3 m (1+ \eta + 3 m \eta)}{(1+3m)(4+3m\eta)}\right].
	\label{eq:SchwartzPlanar}
 \end{equation}
If the limit of vanishing uniform film thickness is not considered, 
the resulting equations for $\eta$ would depend on $H_{\infty}/R$ 
\cite{de2002effect}. 
In the limit of $m\rightarrow 0$, the classical Landau-Levich-Derjaguin 
equation \cite{derjaguin1943,landaulevich} is retrieved for both 
configurations.
Following Bretherton \cite{bretherton1961motion}, Eqs.\ \eqref{eq:SchwartzAxi} 
and \eqref{eq:SchwartzPlanar} can be integrated to find the uniform film 
thickness $H_{\infty}/R$ (see also Cantat \cite{cantat2013liquid} for more 
details). 

\begin{figure}[ht!]
\includegraphics[width=0.45\textwidth]{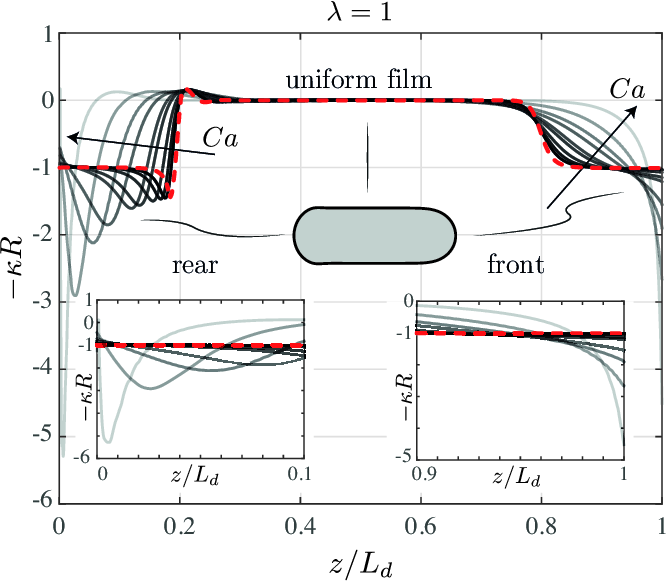}
\caption{Plane curvature of the droplet interface for exponentially increasing
capillary numbers in the range
$10^{-4} < Ca < 1$, obtained 
from FEM-ALE numerical simulations where $\lambda = 1$. The 
dashed red line is for the smallest $Ca$. The $z$ axis is rescaled by the 
droplet 
length to facilitate the comparison. Insets: zoom-in on the 
front and rear menisci. Similar profiles are obtained 
for other viscosity ratios. \label{fig:curvatureEvolution}}
\end{figure}

First, the equations can be linearized in the uniform film region around 
$\eta \approx 1$, giving
\begin{equation}
	\frac{d^3 \eta}{d\xi^3} = K (\eta - 1) ,
	\label{eq:SchwartLin}
\end{equation}
where $K$ is a constant depending on the geometrical 
configurations and the viscosity ratio $m$. Equation \eqref{eq:SchwartLin} has a 
monotonically increasing solution with respect to $\xi$ for the front dynamic 
meniscus, 
$\xi \rightarrow \infty$, and an oscillatorily increasing solution for the 
rear, $\xi \rightarrow -\infty$, as derived in Sec.\ \ref{tx:minFilmThickness}. 
The solution of the front meniscus is $\eta (\xi) = 1 + \alpha 
\exp(K^{1/3} 
\xi)$, where $\alpha$ is a small 
parameter, typically $10^{-6}$. Second, the nonlinear equations \eqref{eq:SchwartzAxi} 
and \eqref{eq:SchwartzPlanar} can be integrated numerically as an initial value problem with a fourth-order 
Runge-Kutta scheme, 
starting from the linear solution until the plane curvature of the interface profile 
becomes constant. A region of constant plane curvature, called static meniscus region (see Fig.~\ref{fig:geometryZones}) exists as 
$d^3\eta/d\xi^3 \approx 0$ for $\eta \gg 1$ (see red line on 
Fig.~\ref{fig:curvatureEvolution}). In the static meniscus region, the interface profile is a parabola: $\eta = P\xi^2/2 + 
C \xi + D$, or, in terms of film thickness,  $H=P(3 Ca)^{2/3} z^2  
/(2H_{\infty})+ C(3Ca)^{1/3}z + D H_{\infty}$, where $P$, $C$ and $D$ 
are real-valued constants. Thus, $P$ is set by the constant plane curvature obtained by the integration of the nonlinear equation.

\begin{figure}[ht!]
\includegraphics[width=0.45\textwidth]{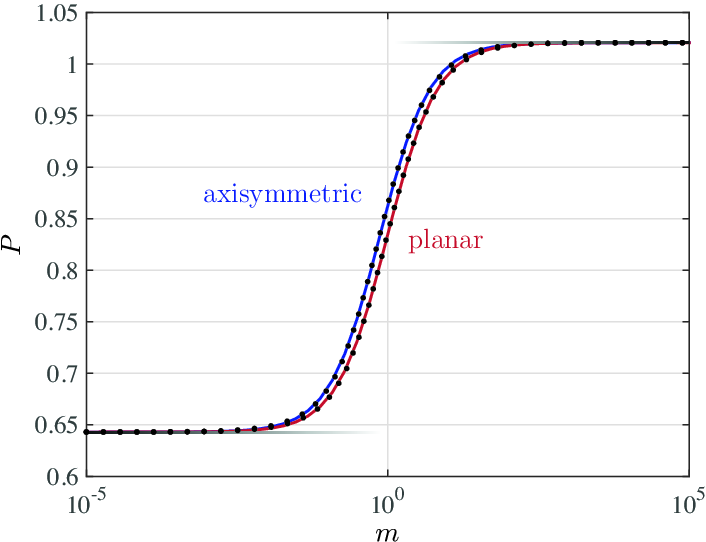}
\caption{Film-thickness coefficient $P$ obtained for discrete $m$ values 
(symbols) and fitting law 
\eqref{eq:Pm} 
(solid lines) as a function of the rescaled viscosity ratio $m$. 
\label{fig:Pm}}
\end{figure}

The procedure can be repeated for any rescaled viscosity ratio $m$ and the 
obtained 
results for the coefficient $P$ can well described by the fitting law 
\cite{schwartz1986motion}:
\begin{equation}
	P(m) =  \frac{0.643}{2}\left\{1 + 2^{2/3} +(2^{2/3}-1)\tanh \left[1.2 \log_{10} m + c_1\right]\right\}
	\label{eq:Pm}
\end{equation}
where the constant $c_1=0.1657$ for the axisymmetric configuration and $c_1= 
0.0159$
for the planar configuration (see Fig.\ \ref{fig:Pm}). The well known limits 
for a bubble $P(0) = 0.643$~\cite{bretherton1961motion} and a very 
viscous droplet 
$P(m\rightarrow \infty) = 2^{2/3} P(0) $ \cite{cantat2013liquid} 
are recovered.

To obtain the uniform film thickness, the matching principle
proposed by Bretherton \cite{bretherton1961motion} is employed. The 
plane curvature $\kappa = d^2H/dz^2 = P(3 Ca)^{2/3} 
/H_{\infty}$ in the static region has to match that of the front 
hemispherical cap of radius $R$, 
which exists for small capillary numbers (see red dashed line in
Fig.~\ref{fig:curvatureEvolution}).
A rigorous asymptotic matching can be found in Park \& Homsy \cite{park1984two}
for a bubble with $m=0$. When $m\neq 0$, the coefficient $P(m)$ depends 
implicitly on 
$H_{\infty}$, and thus on $Ca$, through $m$, leading to an implicit 
asymptotic relation for $H_{\infty}/R$ as:
\begin{equation}
	\frac{H_{\infty}}{R}=P(m) (3 Ca)^{2/3}. 
	\label{eq:HinfImplicit}
\end{equation}
Strictly speaking, the uniform film thickness of viscous droplets ($\lambda \neq 
0$) in the low $Ca$ limit does not scale with $Ca^{2/3}$ as for a 
bubble ($\lambda=0$).

\subsection{Empirical model in the \textit{low}-$Ca$ limit 
\label{tx:filmThicknessSmallCaExplicit}}

\begin{figure}[ht!]
\subfigure[\label{fig:Ca_P_Axis}]{  
\includegraphics[width=0.45\textwidth]{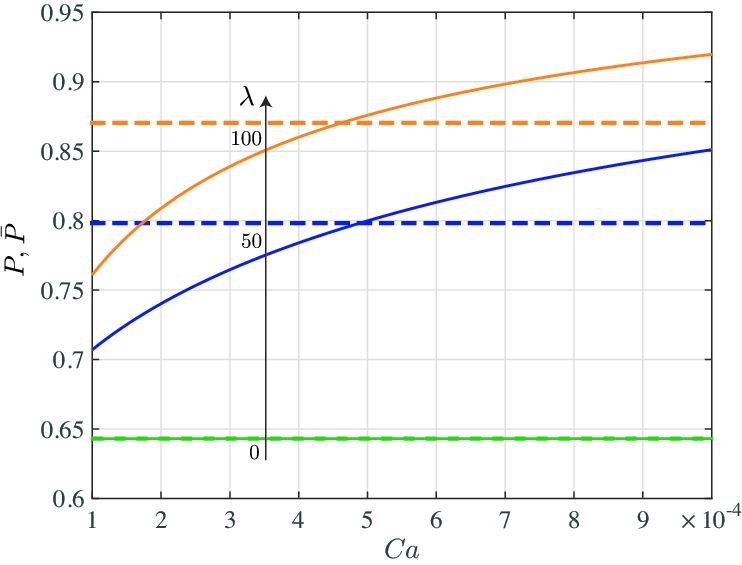}}
\subfigure[\label{fig:Ca_HinfP_Axis}]{  
\includegraphics[width=0.45\textwidth]{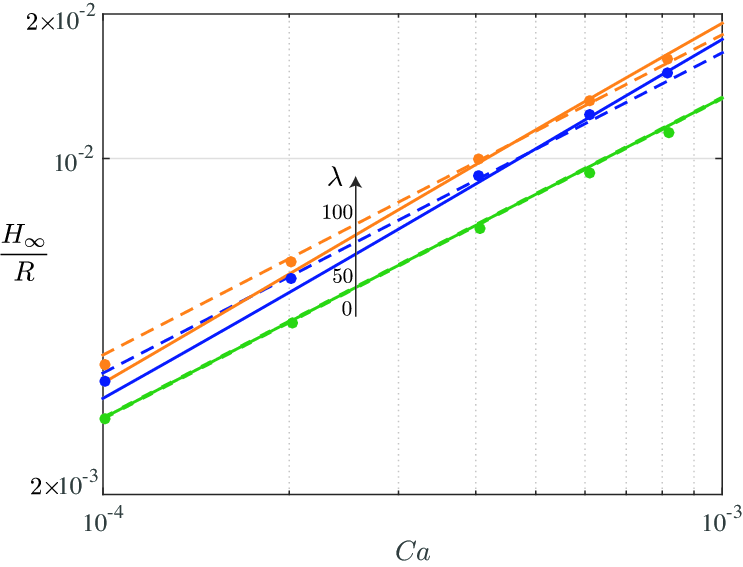}}
\caption{(a) Coefficient $P$ (solid lines) obtained by solving Eq.\ 
\eqref{eq:HinfImplicit} and the mean coefficient 
$\bar{P}$ (dashed lines) for the axisymmetric configuration. The 
viscosity ratios are $\lambda = 0$, $50$ 
and $100$. (b) The uniform film 
thickness $H_{\infty}/R$ from Eq.\ \eqref{eq:HinfImplicit} (solid lines) and Eq.\ 
\eqref{eq:HinfExplicit} (dashed lines), compared to the FEM-ALE simulation 
results (symbols).}
\label{fig:PevolutionsAxis}
\end{figure}

\begin{figure}[ht!]
\includegraphics[width=0.45\textwidth]{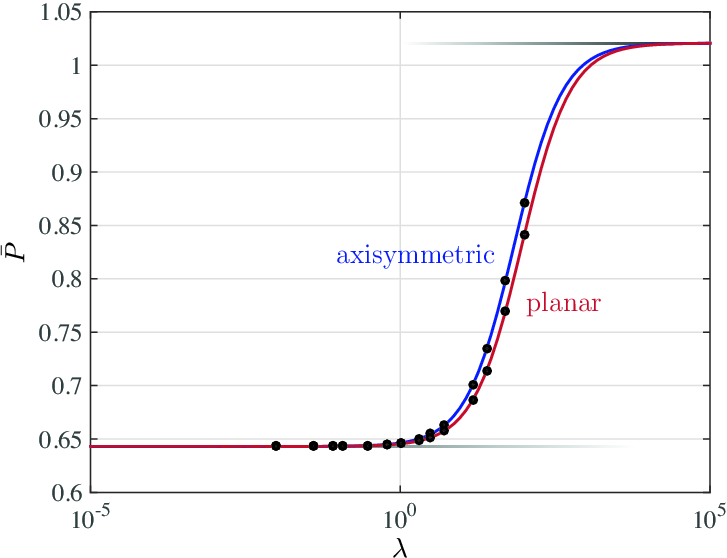}
\caption{Mean coefficient $\bar{P}$ (symbols) obtained by $Ca$-averaging the 
results of the implicit relation Eq.\ \eqref{eq:HinfImplicit} and fitting law (solid 
lines) Eq.\ \eqref{eq:lambda_Pmean} versus the 
viscosity ratio $\lambda$. \label{fig:lambda_Pmean}}
\end{figure}

Equation \eqref{eq:HinfImplicit} holds 
for capillary numbers as low as below $10^{-3}$
\cite{bretherton1961motion}. We solve Eq.\ \eqref{eq:HinfImplicit} numerically and 
present the  coefficient $P$ and 
the uniform film thickness $H_{\infty}/R$ versus $Ca$  
in Fig.\ \ref{fig:PevolutionsAxis} for the axisymmetric case. 
In order to derive an explicit formulation to predict the film
thickness in this $Ca$ regime, we define
$\bar{P}$  as a $Ca$-averaged value of $P$ and define the empirical model
\begin{align}
	\frac{H_{\infty}}{R} =\bar{P}(\lambda) (3 Ca)^{2/3},
	\label{eq:HinfExplicit}
\end{align}
where $\bar{P}(\lambda)$ is independent of $Ca$ (see dashed lines in Fig.\ 
\ref{fig:Ca_P_Axis}) and can be approximated by the fitting law (see Fig.~\ref{fig:lambda_Pmean}):
\begin{align}
	\bar{P} (\lambda) =  \frac{0.643}{2}\{1 + 2^{2/3} 
\nonumber \\
\qquad +(2^{2/3}-1)\tanh \left[1.28 \log_{10} \lambda + 
c_2\right]\}.
	\label{eq:lambda_Pmean}
\end{align}
where the constant $c_2= -2.36$ for the axisymmetric case and 
$c_2= -2.52$ for the planar case are obtained by fitting. For $\lambda = 
0$, $\bar{P}=0.643$ is recovered and $H_{\infty}/R$ indeed scales with 
$Ca^{2/3}$, at least when $Ca<10^{-3}$.
Figure \ref{fig:Ca_HinfP_Axis} also shows that the empirically obtained film 
thickness (dashed lines) Eq.\ \eqref{eq:HinfExplicit} agrees reasonably well 
with the FEM-ALE simulation results (symbols), whereas the implicit 
law (solid lines) Eq.\ \eqref{eq:HinfImplicit} slightly underestimates them at 
very low $Ca$. To cure this mismatch, Hodges et al.~\cite{hodges2004motion} 
proposed a modified interface condition, which however is found to  
overestimate the thickness more than that underestimated by the original 
implicit law.

\subsection{Model for $10^{-3}\lessapprox Ca \lessapprox 1$ \label{tx:filmThicknessLargerCa}}
 
\begin{figure}[ht!]
\subfigure[\label{fig:FittingHinf_lambda1}]{
\includegraphics[width=0.45\textwidth]{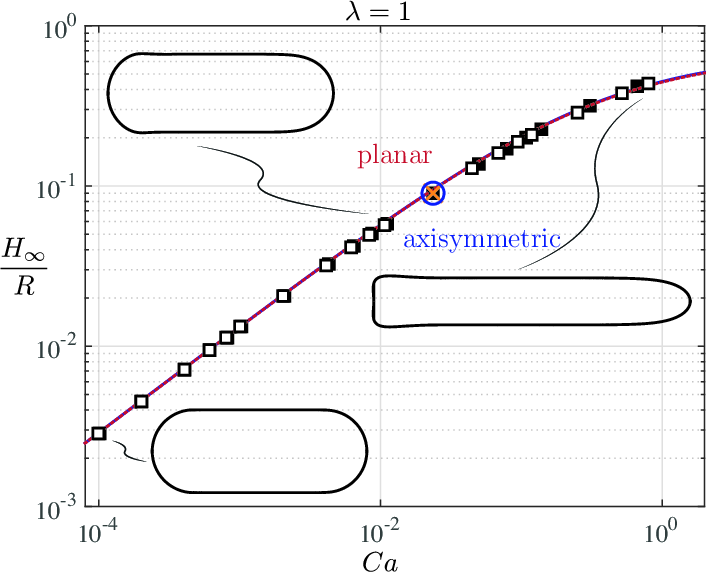}}
\caption{Uniform film thickness given by Eq.\ \eqref{eq:HinfRational} (lines) 
and FEM-ALE numerical results (symbols) as a function of the droplet capillary number 
for $\lambda = 1$ and both axisymmetric (blue solid line, full 
symbols) and planar (dashed red line, empty symbols) geometries. Cross and 
circle correspond to a droplet with $37\%$ and $82\%$, respectively, larger 
volume than the standard one used for the axisymmetric geometry. 
\label{fig:FittingHinf_lambda}}
\end{figure}

Despite the explicit law for the uniform-film thickness prediction with 
$\bar{P}$ proved to be satisfactory, its validity range is restricted to low 
capillary numbers. As known since the experiments of Taylor 
\cite{taylor1961deposition}, the film thickness of a bubble saturates for 
increasing $Ca$. Aussillous and Qu\'er\'e \cite{aussillous2000quick} proposed a model for $\lambda=0$, which agrees well with the experimental data 
of
\cite{taylor1961deposition}, further inspiring the 
two very recent works of Refs.\
\cite{klaseboer2014extended,cherukumudi2015prediction}. In the same vain, we  
propose an empirical model for the film thickness $H_{\infty}$ as a 
function of both $Ca$ and $\lambda$
\begin{equation}
\frac{H_{\infty}}{R}=\frac{\bar{P}(\lambda)(3Ca)^{2/3}}{1+\bar{P}
(\lambda)Q(\lambda)(3Ca)^{2/3}},
	\label{eq:HinfRational}
 \end{equation}
where the coefficient $Q$ is obtained by fitting
Eq.\ \eqref{eq:HinfRational} to the database constructed from our extensive
FEM-ALE simulations over a broad range of $Ca$ for different $\lambda$.
 The proposed function of $Q(\lambda)$ is given in Appendix 
\ref{tx:fittingLaw} and plotted in Fig.~\ref{fig:lambda_Q}. For an axisymmetric 
bubble, we find $Q=2.48$, in accordance with the estimation $Q=2.5$ of Ref.\
\cite{aussillous2000quick}. We now present in 
Fig.~\ref{fig:FittingHinf_lambda} the numerical film thickness  
(symbols) and the empirical model (lines) for $\lambda=1$. For the sake of clarity, the results for $\lambda = 0$ and $100$ are shown in the appendix \ref{tx:additionalResults} on Fig.~\ref{fig:FittingHinf_lambda_appendix}.
For $\lambda = 1$, the thickness of the two configurations coincide. However, when $Ca \sim O(1)$, the film is thicker in the planar 
configuration than in the axisymmetric one for a bubble ($\lambda=0$); the trend
reverses for a highly-viscous droplet ($\lambda=100$). This $\lambda$-dependence  
of the film thickness is indeed implied by the crossover of the two 
fitting functions $Q(\lambda)$ at $\lambda = 1 $ shown in Fig.~\ref{fig:lambda_Q}.

\begin{figure}[h!]
\includegraphics[width=0.45\textwidth]{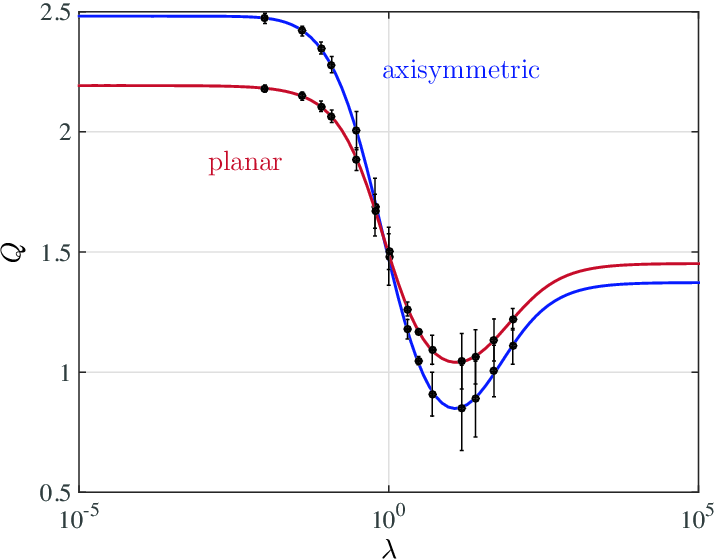}
\caption{Coefficient $Q$ obtained for the simulated viscosity ratios (symbols) and 
proposed fitting law (see Appendix \ref{tx:fittingLaw}) as a function of the 
viscosity ratio $\lambda$. \label{fig:lambda_Q}}
\end{figure}

It has to be noted that when the capillary number is increased, 
the regions of constant plane curvature in the static front and rear caps reduce in 
size and eventually disappear (see Fig.\ \ref{fig:curvatureEvolution}), and this 
for all viscosity ratios. The matching to a region of constant plane curvature for 
large capillary numbers as proposed by 
Refs.~\cite{klaseboer2014extended,cherukumudi2015prediction} might be 
questionable for this $Ca$-range.

The uniform film thickness of droplets with $37\%$ and $82\%$ larger volume, resulting in longer droplets, are 
compared on Fig.~\ref{fig:FittingHinf_lambda}, showing that as long as such a 
uniform region exists, the results are independent of the droplet length.

\section{Droplet velocity 
\label{tx:dropletVelocityUinf}}

Equipped with the model of the uniform-film thickness $H_{\infty}$, 
we derive the droplet velocity based on the velocity profiles in the 
uniform-film region given in Sec.\ \ref{tx:flowProfile}. 
At the location $H=H_{\infty}$ where the 
interface is flat, the pressure gradients are equal, $dp_i/dz=dp_o/dz=dp/dz$. 
We further use $q_o= \pi R^2 (U_{\infty}-U_d)$ 
imposed by mass conservation and $q_i=0$ (in the moving frame 
of the droplet) to obtain the analytical expressions for the pressure gradient 
\begin{equation}
	\frac{d p}{dz} \bigg\vert_{r=R-H_{\infty}} =   \frac{-8 R^2 U_{\infty} 
\mu_i}{(R-H_{\infty})^4+H_{\infty}(2R-H_{\infty})(2R^2-2H_{\infty}R 
+H_{\infty}^2)\lambda}, 
	\label{eq:dpdzParallelAxi}
\end{equation}
 and the droplet velocity
\begin{equation}
	U_d=  \frac{R^2[(R-H_{\infty})^2+2 H_{\infty}(2R-H_{\infty})\lambda]}{(R-H_{\infty})^4+H_{\infty}(2R-H_{\infty})(2R^2-2H_{\infty}R+H_{\infty}^2)\lambda}U_{\infty} .
	\label{eq:Ud}
\end{equation}
The relative velocity  of the 
axisymmetric droplet with respect to the underlying velocity reads
\begin{equation}
	\frac{U_d-U_{\infty}}{U_d} = \frac{\left(2-\frac{H_{\infty}}{R}\right)\frac{H_{\infty}}{R} \left[ 1+ \left(2-\frac{H_{\infty}}{R}\right)\frac{H_{\infty}}{R}(\lambda-1) \right] }{1+\left(2-\frac{H_{\infty}}{R}\right) \frac{H_{\infty}}{R}(2\lambda -1)} .
	\label{eq:relVelAxis}
\end{equation}
An analogous derivation for the planar configuration yields (see Appendix \ref{tx:velocityPlanar}):
\begin{equation}
	\frac{U_d-U_{\infty}}{U_d} = \frac{\frac{H_{\infty}}{R} \left\{2-\frac{H_{\infty}}{R} \left[4+2 \frac{H_{\infty}}{R}(\lambda-1) -3 \lambda\right] \right\}}{2+\left(2-\frac{H_{\infty}}{R}\right)\frac{H_{\infty}}{R}(3\lambda -2)}.
	\label{eq:relVelPlanar}
\end{equation}

\begin{figure}[ht!]
\subfigure[\label{fig:DropletVelocity_lambda0}]{
\includegraphics[width=0.45\textwidth]{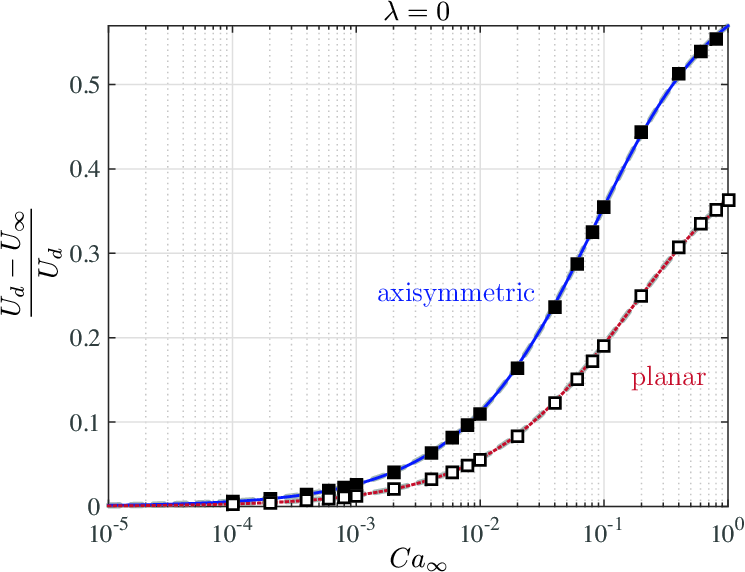}}
\subfigure[\label{fig:DropletVelocity_lambda1}]{
\includegraphics[width=0.45\textwidth]{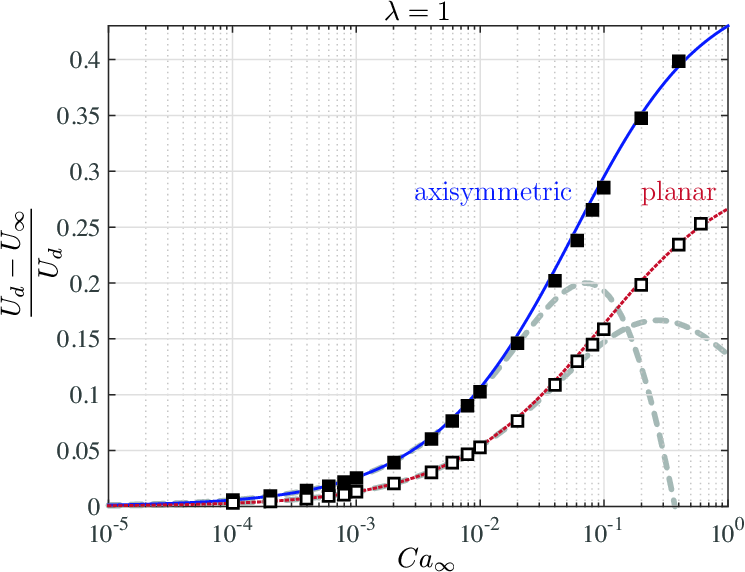}}
\subfigure[\label{fig:DropletVelocity_lambda100}]{
\includegraphics[width=0.45\textwidth]{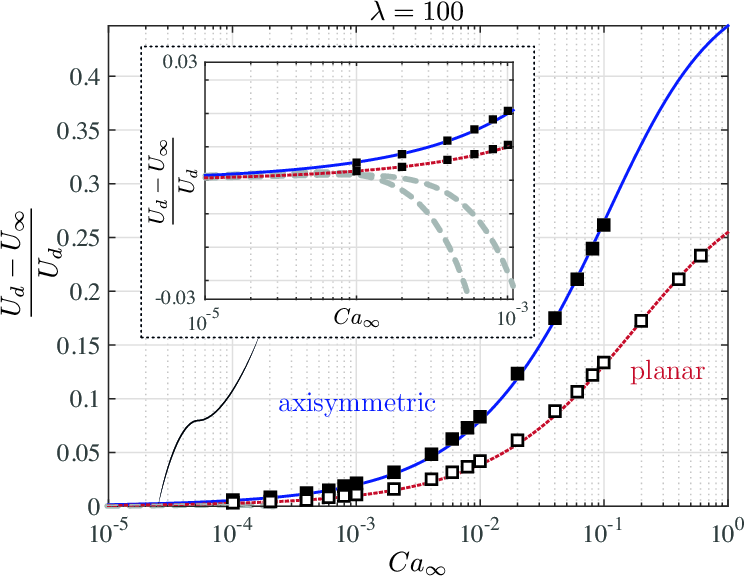}}
\caption{Relative droplet velocity  (lines)  predicted by Eqs.\ \eqref{eq:relVelAxis} and 
\eqref{eq:relVelPlanar} together with the proposed empirical model for the uniform film 
thickness \eqref{eq:HinfRational} and the results of the FEM-ALE 
numerical 
simulations (symbols) as a function of capillary number $Ca_{\infty}$  
for $\lambda = 0$ (a), $1$ (b) and $100$ (c) and both axisymmetric (blue solid line, full 
symbols) and planar (dashed red line, empty symbols) geometries. Long dashed 
gray lines correspond to the asymptotic estimates of Eqs.\ \eqref{eq:relVelAxiAsympt} and \eqref{eq:relVelPlanarAsympt}. \label{fig:dropletVelocity}}
\end{figure}

Eqs. \eqref{eq:HinfRational} and \eqref{eq:Ud} form a system of the two 
unknowns, namely the droplet capillary number $Ca$ and the uniform film 
thickness $H_{\infty}/R$. It is important to remind that the former is related to the droplet 
velocity via $Ca = Ca_{\infty}U_d/U_{\infty}$. For a given 
combination of inflow capillary number $Ca_{\infty}$ and viscosity ratio $\lambda$ as the input, the system can be 
solved numerically (see Matlab file \texttt{filmThicknessAndVelocity.m} in the Supplementary Material) 
outputting $Ca$ and $H_{\infty}/R$. The predicted relative velocity 
$({U_d-U_{\infty}})/{U_d}$ (lines) agrees well the FEM-ALE simulation 
results (symbols) as shown in
Fig.\ 
\ref{fig:dropletVelocity}.

In the limit of ${H_{\infty}}/{R}\rightarrow 0$, the relative velocity can 
be approximated asymptotically as 
\begin{equation}
\frac{U_d-U_{\infty}}{U_d} =  2 \left(\frac{H_{\infty}}{R}\right) - (1+4 
\lambda) \left(\frac{H_{\infty}}{R}\right)^2 + 
O\left(\frac{H_{\infty}}{R}\right)^3
\label{eq:relVelAxiAsympt}
\end{equation}
for the axisymmetric case, and
\begin{equation}
\frac{U_d-U_{\infty}}{U_d} =    
\left(\frac{H_{\infty}}{R}\right)-\frac{3\lambda}{2} 
\left(\frac{H_{\infty}}{R}\right)^2 + O\left(\frac{H_{\infty}}{R}\right)^3
\label{eq:relVelPlanarAsympt}
\end{equation}
for the planar geometry.
For very low capillary numbers, the asymptotic 
estimates predict that the relative droplet velocity scales with 
$H_{\infty}/R$, and hence with $Ca^{2/3}$ \cite{stone2010interfaces}.
The viscosity ratio $\lambda$ only enters at second order of $H_{\infty}/R$,
which however influences the validity 
range of the asymptotic estimates \eqref{eq:relVelAxiAsympt} and 
\eqref{eq:relVelPlanarAsympt}  considerably. The asymptotic estimates are exact 
for $\lambda = 0$. 
In this case, Eqs.\ \eqref{eq:relVelAxiAsympt} and \eqref{eq:relVelPlanarAsympt} 
reduce to the well known predictions for bubbles 
$(2-H_{\infty}/R)H_{\infty}/R$ and $H_{\infty}/R$ \cite{bretherton1961motion,stone2010interfaces,langewisch2015prediction}, respectively (see Fig.\ 
\ref{fig:DropletVelocity_lambda0}). For non-vanishing $\lambda$, the 
complete expressions \eqref{eq:relVelAxis} and \eqref{eq:relVelPlanar} should be 
employed (see Fig.\ \ref{fig:DropletVelocity_lambda1}). For example, the 
asymptotic estimate for $\lambda = 100$ is only valid when $Ca_{\infty} < 
10^{-4}$ (see Fig.\ \ref{fig:DropletVelocity_lambda100}).

\section{Minimum film thickness \label{tx:minFilmThickness}}

\begin{figure}[ht!]
\subfigure[\label{fig:Gm}]{
\includegraphics[width=0.45\textwidth]{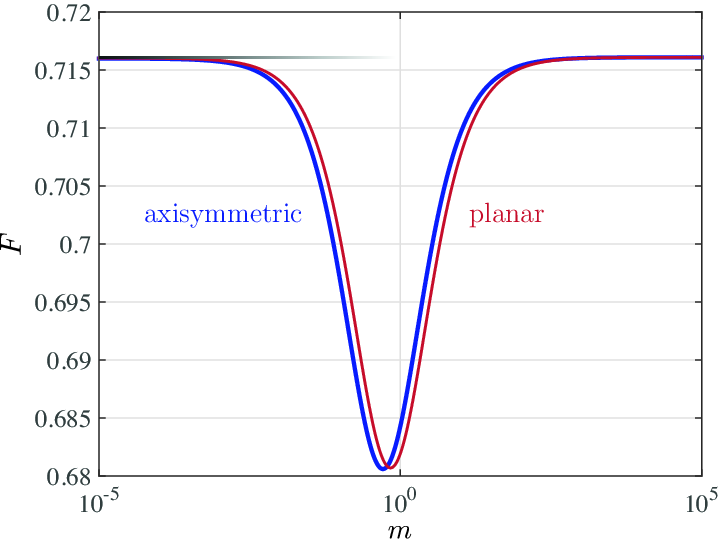}}
\subfigure[\label{fig:lambda_Gmean}]{
\includegraphics[width=0.45\textwidth]{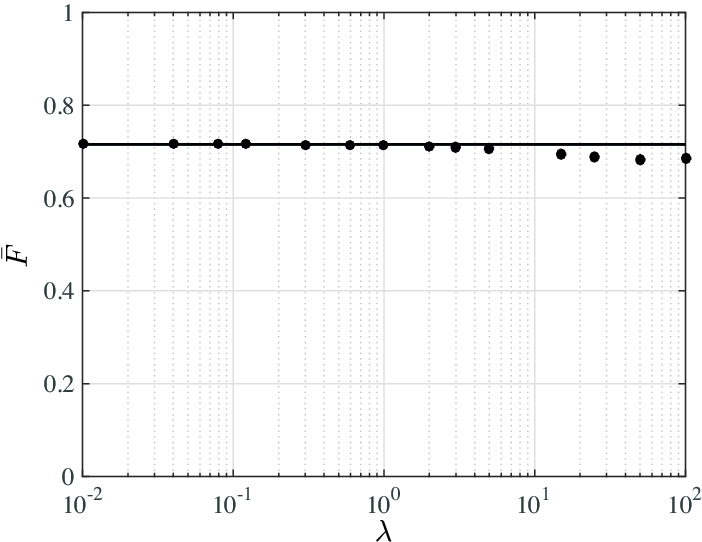}}
\caption{(a) Minimum film-thickness coefficient $F$ as a function
of the rescaled viscosity ratio $m$. (b) Mean coefficient $\bar{F}$ (symbols) and 
fitting law (solid line) as a function of the viscosity ratio $\lambda$ 
\label{fig:G}}
\end{figure}

At low capillary numbers $Ca$, the droplet interface exhibits an oscillatory 
profile between the uniform thin film and the rear static cap (see
Fig.\ \ref{fig:curvatureEvolution}). The minimum film thickness in the low $Ca$ 
limit can be computed by integrating 
the lubrication equation \eqref{eq:SchwartzAxi} or \eqref{eq:SchwartzPlanar} 
for $\xi =0 $ to $\xi \rightarrow -\infty$. The initial condition for this initial value problem is given by the solution of the linear equation \eqref{eq:SchwartLin} for negative $\xi$: $\eta = 1+\alpha \exp (-K^{1/3} \xi/2) 
\cos(\sqrt{3} K^{1/3} \xi/2+\phi)$, where $\alpha$ is a small parameter of  
order $10^{-6}$ and $\phi$ is a parameter taken such that the constant 
plane curvature of the nonlinear integrated solution at $\xi \rightarrow -\infty$ is 
equal to the one of the front static cap 
\cite{bretherton1961motion,cantat2013liquid} as discussed in Sec.\ 
\ref{tx:filmThicknessSmallCaImplicit}. Note that the linear solution for the rear dynamic meniscus presents oscillations.
The minimum film thickness of the obtained profile is found to follow the empirical model 
\cite{bretherton1961motion}
\begin{equation}
	\frac{H_{\text{min}}}{R} = F(m) P(m) (3Ca)^{2/3}  \quad \text{with} \quad m=\lambda \frac{H_{\infty}}{R} ,
	\label{eq:G}
\end{equation}
where $F(m)$ is a coefficient obtained through fitting Eq.~\eqref{eq:G} 
to our numerical database (see Fig.\ \ref{fig:Gm}).
Similar to the mean coefficient $\bar{P}$ adopted in Sec.\ 
\ref{tx:filmThickness}, a $Ca$-averaged $F(m)$ can be introduced as
$\bar{F}$, that is further assumed  as $0.716$ in view of its very weak 
dependence on $\lambda$ shown in Fig.\ \ref{fig:lambda_Gmean}.

The minimum film thickness is bounded by the thickness of the uniform 
film and hence will saturate at large capillary numbers. Thus, for 
sufficiently large $Ca$ values, the oscillations at the rear interface would
disappear and 
$H_{\text{min}}=H_{\infty}$. It is therefore natural to 
propose a rational function model of $H_{\text{min}}$ for a 
broader $Ca$-range as the one for $H_{\infty}$:
 \begin{equation}
 	\frac{H_{\text{min}}}{R}=\frac{\bar{P}(\lambda)\bar{F}(3Ca)^{2/3}}{1+\bar{P}(\lambda)\bar{F}G(\lambda)(3Ca)^{2/3}} .
	\label{eq:HminRational}
 \end{equation}

\begin{figure}[ht!]
\subfigure[\label{fig:FittingHmin_lambda1}]{
\includegraphics[width=0.45\textwidth]{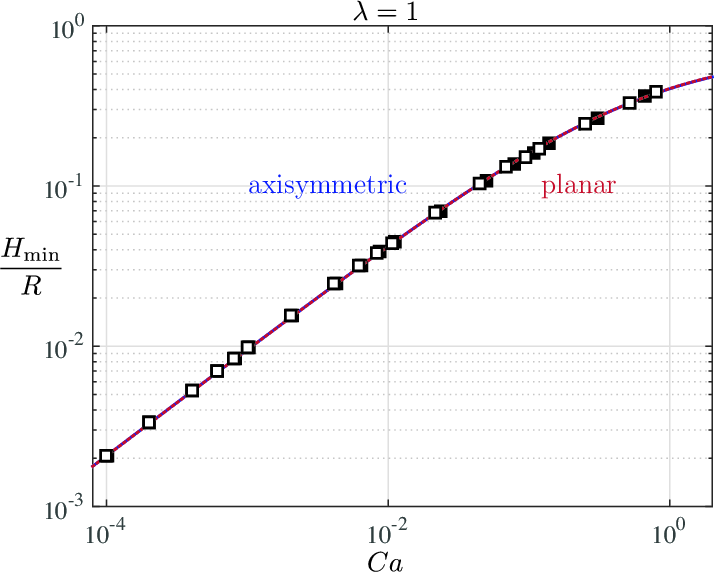}}
\caption{Minimum film thickness given by Eq.\ \eqref{eq:HminRational} (lines) 
and FEM-ALE numerical results (symbols) as a function of the droplet capillary 
number for $\lambda = 1$ and both axisymmetric (blue 
solid line, full symbols) and planar (dashed red line, empty symbols) 
geometries. \label{fig:FittingHmin_lambda}}
\end{figure}

The above  minimum film thickness model \eqref{eq:HminRational} 
together with the coefficient $G$ is in good agreement with the results of the 
numerical simulations (see Fig.\ \ref{fig:FittingHmin_lambda} and Fig.\ \ref{fig:FittingHmin_lambda_appendix}).
The proposed fitting of the coefficient $G$ as a function of the viscosity 
ratio (see Fig.\ \ref{fig:lambda_F}) is given in Appendix \ref{tx:fittingLaw}.
\begin{figure}[ht!]
\includegraphics[width=0.45\textwidth]{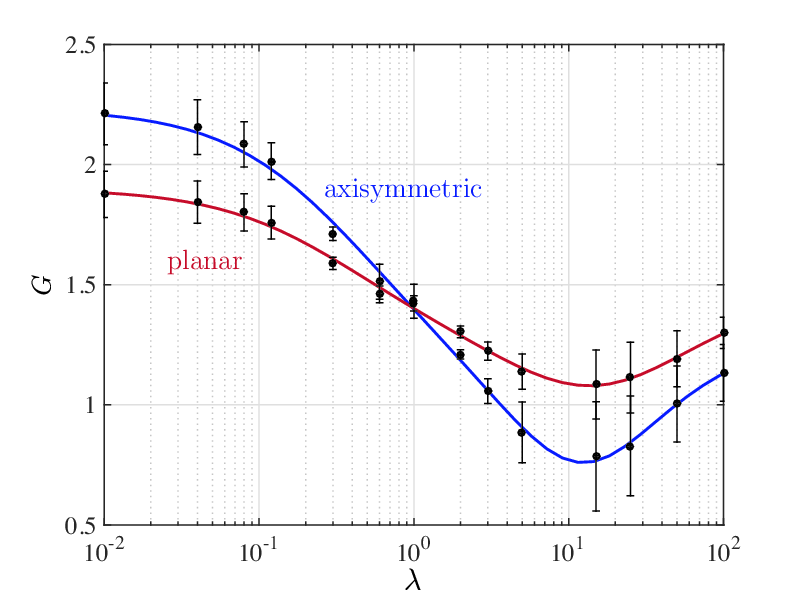}
\caption{Coefficient $G$ obtained for the simulated viscosity ratios (symbols) and 
proposed fitting law \eqref{eq:fittingLaw} (solid lines) as a function of the 
viscosity ratio $\lambda$. \label{fig:lambda_F}}
\end{figure}

\section{Front and rear total stress jumps \label{tx:frontRearStressJump}}

The dynamics of a translating bubble in a capillary tube has been characterized 
since the seminal work of 
Bretherton \cite{bretherton1961motion} not only by the mean and 
minimum film thickness, the relative velocity compared to the mean velocity, but 
also by the mean curvature of the front and rear static menisci. In fact, for 
$Ca\rightarrow0$, the pressure drop across the interface is directly related to the expression of its mean curvature via the Laplace law.
Having generalized the film thickness and droplet velocity models for non-vanishing 
viscosity ratios, we are hereby focusing on the evolution of the 
plane curvature of the front and rear static caps versus the capillary number and 
viscosity ratio. The mean curvature at the droplet extremities is equivalent to the corresponding plane curvature for the planar configuration or to its double for the axisymmetric configuration: $\mathcal{C}_{f,r} = \chi \kappa_{f,r}$, with $\chi=2$ (resp. $\chi=1$) for the axisymmetric (resp. planar) configuration (see Sec.~\ref{tx:governingEquations}). As it will be shown, given the rather broad range of capillary 
numbers considered (approaching $O(1)$), it is
insufficient to consider the interface mean curvature alone to provide an accurate 
prediction of the pressure drop, but the jump in the normal viscous 
stress has to be accounted for.

For the incompressible Newtonian fluids considered, the viscous stress tensor is $\boldsymbol{\tau} =  
\mu \left[ 
\left( \nabla 
\vec{u} \right) + \left( \nabla \vec{u} \right)^T \right]$, and hence the 
$z$-direction normal total stress $\sigma_{zz}$ is given by
\begin{equation}
\sigma_{zz} = - p + \tau_{zz}  = -p + 2 \mu \frac{\partial u}{\partial z}.
\label{eq:sigmazz}
\end{equation}
Applying the difference (between inner and outer phases) operator $\Delta$ to 
Eq.~\eqref{eq:sigmazz} and based on the dynamic boundary condition in the normal direction \eqref{eq:bcNorm} 
at the droplet front and rear extremities, we get
\begin{equation}
	\Delta \sigma_{zz_{f,r}}= - \Delta p_{f,r} +  \Delta \tau_{zz_{f,r}} 
	= - \gamma \chi \kappa_{f,r},
	\label{eq:bcDynCaps}
\end{equation}
which indicates that the total stress jump at the front/rear extremities 
scales with the local interface mean curvature and is 
the sum of the pressure jump and the normal viscous stress jump. 
These quantities will be modeled separately in the following sections.

\subsection{Front and rear plane curvatures \label{tx:frontRearCurvatures}}

\begin{figure}[ht!]
\subfigure[\label{fig:FittingKf_lambda1}]{
\includegraphics[width=0.45\textwidth]{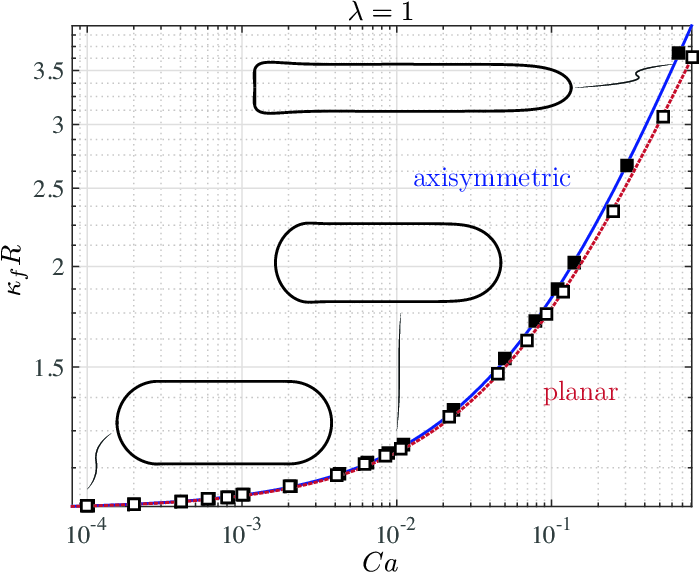}}
\caption{Curvature $\kappa_f$ of the front meniscus predicted by the model Eq.\ \eqref{eq:KfKr} (lines) 
and FEM-ALE data (symbols) versus $Ca$ for both axisymmetric (blue line, full
symbols) and planar (red dashed line, empty symbols) geometries, where the 
viscosity ratio $\lambda = 1$. 
\label{fig:FittingKf_lambda}}
\end{figure}

\begin{figure}[ht!]
\subfigure[\label{fig:FittingKr_lambda1}]{
\includegraphics[width=0.45\textwidth]{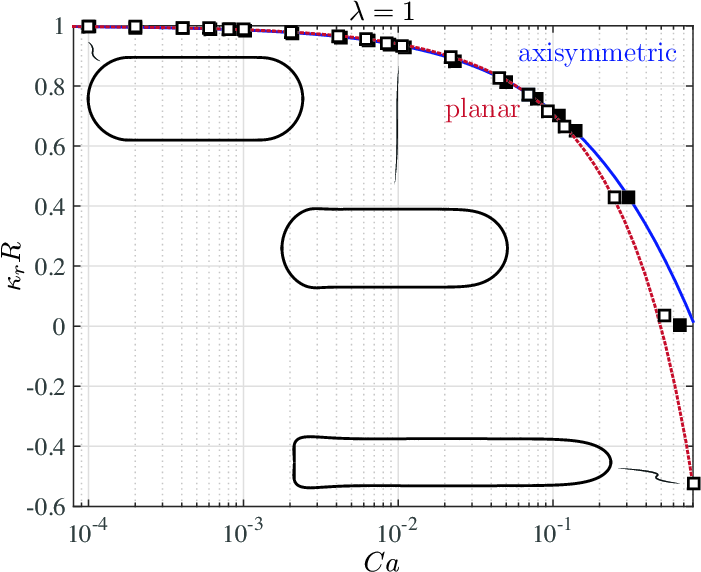}}
\caption{The rear counterpart $\kappa_r$ of Fig.~\ref{fig:FittingKf_lambda}.
\label{fig:FittingKr_lambda}}
\end{figure}

In the spirit of the empirical film thickness model, 
the plane curvature $\kappa_f$ of the front meniscus  and that of the rear, $\kappa_r$, are approximated by the rational function model
\begin{equation}
	\kappa_{f,r}R  = \frac{1+T_{f,r}(\lambda)(3 Ca)^{2/3}}{1+Z_{f,r}(\lambda) (3 Ca)^{2/3}},
	\label{eq:KfKr}
\end{equation}
where $T_{f,r}$ and $Z_{f,r}$ as  
$\lambda$-dependent constants are obtained by fitting Eq.\ \eqref{eq:KfKr} to 
the FEM-ALE data (see Appendix 
\ref{tx:fittingLaw}). It is worth-noting that the asymptotic 
series of the proposed expression,
\begin{equation}
	\kappa_{f,r}R \sim 1+(T_{f,r}-Z_{f,r})(3Ca)^{2/3}+O(Ca^{4/3}),
\end{equation}
is in line with the law proposed by Bretherton 
\cite{bretherton1961motion}, namely $1+\beta_{f,r}(3Ca)^{2/3}+O(Ca^{4/3})$. 
Thus, the empirical model \eqref{eq:KfKr}, which is in excellent agreement with 
the numerical results (see Fig.~\ref{fig:FittingKf_lambda} and 
~\ref{fig:FittingKr_lambda} as well as Fig.~\ref{fig:FittingKf_lambda_appendix} and 
~\ref{fig:FittingKr_lambda_appendix}), can be regarded as an empirical extension of Bretherton's 
law to a broader capillary numbers range up to $1$. The mean curvature at the droplet extremities is given by $\mathcal{C}_{f,r} = \chi \kappa_{f,r}$.

%


\subsection{Front and rear pressure jumps -- classical model 
\label{tx:frontRearPJumps}}

\begin{figure}[ht!]
\subfigure[\label{fig:FittingDpf_lambda0}]{
\includegraphics[width=0.45\textwidth]{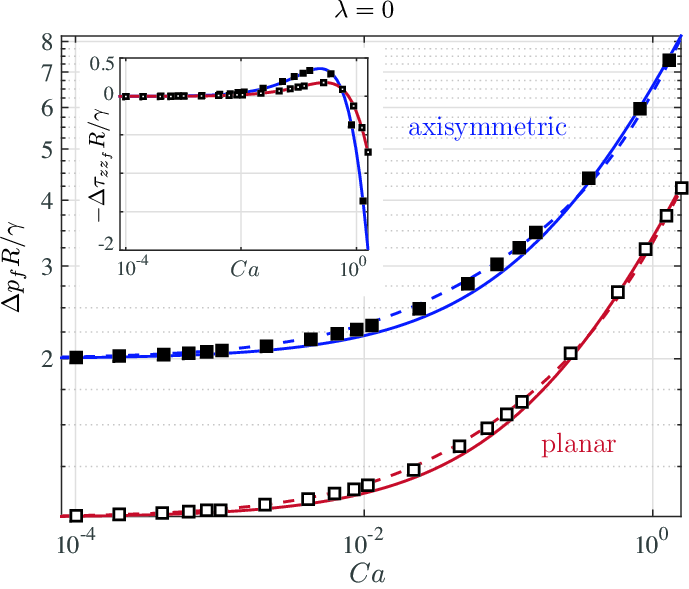}}
\caption{Front pressure jump $\Delta p_f$ given by Eq.\ \eqref{eq:DpfDpr} 
(solid lines) 
and front normal viscous stress jump $\Delta \tau_{{zz}_{f}}$  by Eq.\ 
\eqref{eq:DUzfDUzr} (inset, solid lines) and FEM-ALE data (symbols) versus 
$Ca$ for both axisymmetric (blue line, 
full symbols) and planar (red line, empty symbols) geometries, where the 
viscosity ratio $\lambda = 0$. The dashed lines 
correspond to the improved pressure jump model Eq.\ \eqref{eq:DpfDprImproved}. 
Note the different scale in the insets. \label{fig:FittingDpf_lambda}}
\end{figure}

\begin{figure}[ht!]
\subfigure[\label{fig:FittingDpr_lambda0}]{
\includegraphics[width=0.45\textwidth]{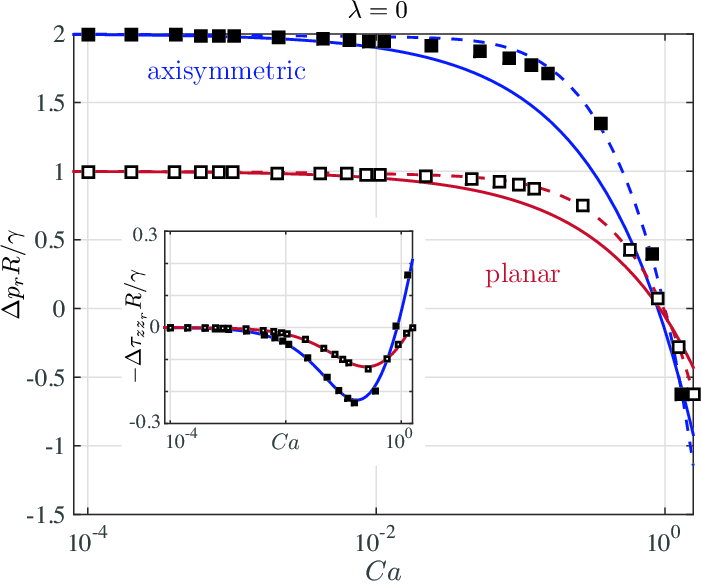}}
\caption{The rear counterpart, pressure jump $\Delta p_r$ 
and normal viscous stress jump $\Delta \tau_{{zz}_{r}}$, of 
Fig.~\ref{fig:FittingDpf_lambda}.
\label{fig:FittingDpr_lambda}}
\end{figure}

Following the literature \cite{bretherton1961motion,cherukumudi2015prediction}, the dimensionless pressure jump 
${\Delta p_{f,r} R}/{\gamma}=(p_{{f,r}}^i - p_{{f,r}}^o)R/\gamma$ at the front 
and rear of the droplet is described by the empirical model
\begin{equation}
	\frac{\Delta p_{f,r} R}{\gamma} = \chi \left[1+S_{f,r}(\lambda) (3 Ca)^{2/3}\right],
	\label{eq:DpfDpr}
\end{equation}
where $\chi=2$ (resp. $\chi=1$) for the axisymmetric (resp. planar), 
and $S_{f,r}$ is a $\lambda$-dependent coefficient. Equation
\eqref{eq:DpfDpr} is in fact inspired by the curvature model 
proposed by Bretherton \cite{bretherton1961motion} exploiting the Laplace law 
\cite{cherukumudi2015prediction}, reason why we call it classical model.
The coefficient $S_{f,r}$ could be derived from the integration of the lubrication 
equation.~\eqref{eq:SchwartzAxi} or \eqref{eq:SchwartzPlanar}, which is 
valid in the low-$Ca$ 
limit when the viscous stresses and their jumps are 
negligible. To broaden the $Ca$ range of the model, we obtain 
$S_{f,r}$ through fitting to the FEM-ALE data. Nevertheless,
 as visible in Fig.\ 
\ref{fig:FittingDpf_lambda} and 
Fig.\ \ref{fig:FittingDpr_lambda}, the model fails to precisely describe
the numerical data, particularly for the rear pressure jump 
at high $Ca$ values (see Fig.\ 
\ref{fig:FittingDpr_lambda}).

After explaining our model for the normal viscous stress jump in 
Sec.~\ref{tx:frontRearUzJumps}, we will show in Sec.\ 
\ref{tx:frontRearPJumpsImproved} 
that the pressure jump can be better approximated by summing up the 
two contributions from the interface mean curvature and the normal viscous stress 
jump, which are modeled separately. The importance of the normal viscous stress 
jump for the pressure jump is already noticeable when comparing the 
evolutions of 
the plane curvature $\kappa_{f,r}$ and the one of the pressure jump $\Delta p_{f,r} 
/ \gamma$ in Figs.\ \ref{fig:FittingKf_lambda} and \ref{fig:FittingDpf_lambda} 
or in Figs.\ \ref{fig:FittingKr_lambda} and \ref{fig:FittingDpr_lambda}. 


\subsection{Front and rear normal viscous stress jumps \label{tx:frontRearUzJumps}}

The dimensionless normal viscous stress jump
$\Delta \tau_{zz} R /\gamma= (\tau_{{zz}_{f,r}}^i - \tau_{{zz}_{f,r}}^o 
)R/\gamma$ at the front and rear of the droplet is approximated by the 
following model
\begin{equation}
	\frac{\Delta \tau_{zz_{f,r}} R}{\gamma} = \frac{M_{f,r}(\lambda) 
(3Ca)+N_{f,r}(\lambda) (3Ca)^{4/3}}{1+O_{f,r}(\lambda)(3Ca)},
	\label{eq:DUzfDUzr}
\end{equation}
where $M_{f,r}$, $N_{f,r}$ and $O_{f,r}$ are viscosity ratio dependent
coefficients found by fitting Eq.\ \eqref{eq:DUzfDUzr} to the FEM-ALE data. The 
 normal viscous stress jumps  indeed scale with $Ca$ for small capillary 
numbers, as found by Bretherton \cite{bretherton1961motion}. The 
comparison between the model and the numerical results is 
shown in the insets of Figs.\ \ref{fig:FittingDpf_lambda} and 
\ref{fig:FittingDpr_lambda}, where the results for $\lambda = 0$ are shown. The results for $\lambda = 1$ and $100$ can be found in Figs.\ \ref{fig:FittingDpf_lambda_appendix} and \ref{fig:FittingDpr_lambda_appendix}. The stress jump $\Delta \tau_{zz}$ is found to be small 
in the case of $\lambda=1$ and it varies with $Ca$ non-monotonically
for the other viscosities.

\subsection{Front and rear pressure jumps  -- improved model \label{tx:frontRearPJumpsImproved}}

Using the dynamic boundary condition in the normal direction evaluated at the front and rear caps of the droplet, Eq.~\eqref{eq:bcDynCaps}, the pressure jump at 
the front and rear caps can also be 
computed as
\begin{equation}
	\Delta p_{f,r}  =  \gamma \chi \kappa_{f,r} +  \Delta \tau_{zz_{f,r}}.
	\label{eq:dpKTau}
\end{equation}
Thus, with the proposed models \eqref{eq:KfKr} and \eqref{eq:DUzfDUzr}
for the interface curvatures and normal viscous stress jumps at hand, the pressure jump model reads
\begin{align}
	\frac{\Delta p_{f,r} R}{\gamma} =& \frac{M_{f,r}(\lambda) (3Ca)+N_{f,r}(\lambda) (3Ca)^{4/3}}{1+O_{f,r}(\lambda)(3Ca)} \nonumber \\
	&+ \chi \frac{1+T_{f,r}(\lambda)(3 Ca)^{2/3}}{1+Z_{f,r}(\lambda) (3 Ca)^{2/3}},
	\label{eq:DpfDprImproved}
\end{align}
which agrees with the FEM-ALE data better than Eq. \eqref{eq:DpfDpr} does
(see dashed lines on Figs.\ \ref{fig:FittingDpf_lambda} and 
\ref{fig:FittingDpr_lambda} or Figs.\ \ref{fig:FittingDpf_lambda_appendix} and 
\ref{fig:FittingDpr_lambda_appendix}). Therefore, the jump in normal viscous stresses 
has to be taken into account for $Ca>10^{-3}$.

\section{Stresses distribution and total pressure drop 
\label{tx:dropletVelocityDp}}

\subsection{Stresses distribution along the channel centerline 
\label{tx:stressesEvolCenterline}}

The flow field can be classified into parallel and non-parallel regions (see 
Fig.\ \ref{fig:streamlines}). The parallel regions compose of the 
region sufficiently far away from the droplet and that encompassing
the uniform lubrication film of constant thickness, where the flow is
streamwise invariant.
The profile of the streamwise velocity $u(r)$ is parabolic
and the radial velocity $v \approx 0$.
On the contrary, the flow  is not parallel near the droplet extremities
(see Fig.\ \ref{fig:streamlines}) where the flow stagnates. 
Therefore the nearby streamwise velocity vary significantly, leading to a 
non-zero radial velocity $v$ owing to the divergence-free 
condition.

\begin{figure}[ht!]
\subfigure[\label{fig:stressEvolution_Lam1_CaInf0.0008_axis}]{
\includegraphics[width=0.45\textwidth]{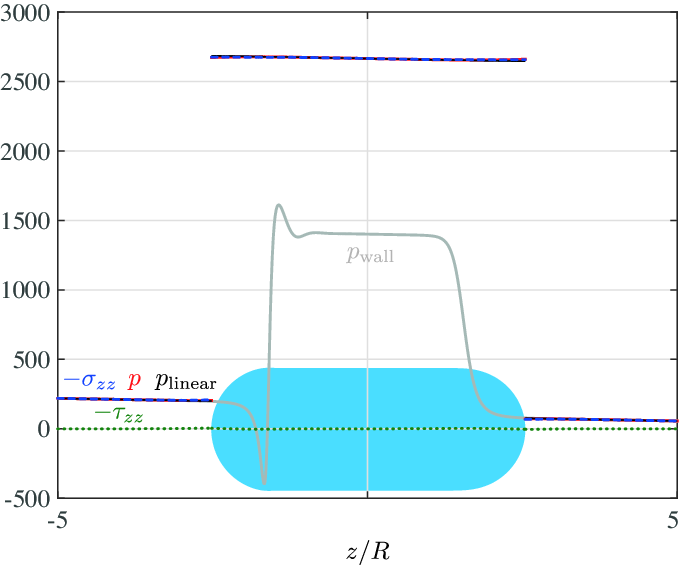}}
\subfigure[\label{fig:stressEvolution_Lam1_CaInf0.008_axis}]{
\includegraphics[width=0.45\textwidth]{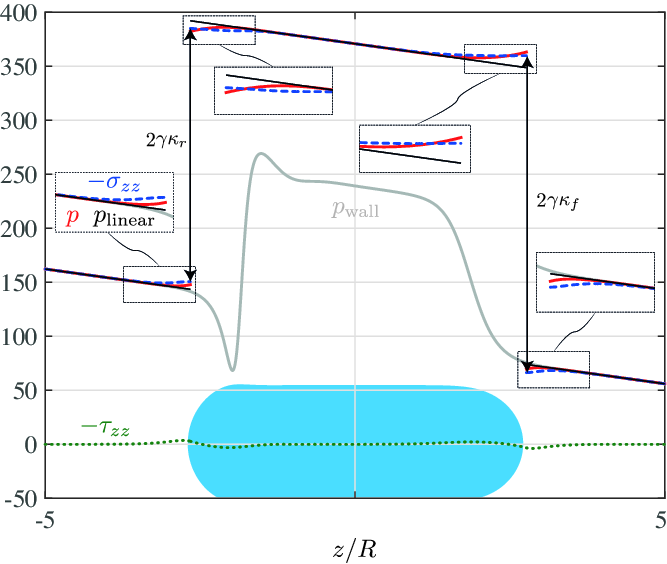}}
\caption{Spatial evolution of the pressure $p$ (red solid line), normal viscous 
stresses
$-\tau_{zz}$ (green dotted 
line) and total stresses $-\sigma_{zz}$ (blue dashed line) along the centerline 
for $Ca = 8.2\cdot 10^{-4}$ (a) and  $Ca = 8.8\cdot 10^{-3}$  (b), $\lambda = 1$ 
and an axisymmetric configuration, obtained from the FEM-ALE numerical simulations. The linear pressure evolution 
without considering non-parallel flow effects is shown by the thin black lines. 
The total stresses jumps induced by the mean curvature 
at the interfaces are indicated by arrows. The droplet shape is indicated 
in blue. The pressure at the channel wall is indicated by the grey line. 
\label{fig:stressEvolutions}}
\end{figure}

We show in Fig.\ \ref{fig:stressEvolutions} the distribution of the 
total stress component
$\sigma_{zz}=-p + \tau_{zz}$, of the pressure $p$ and of the viscous stress component
$\tau_{zz}=2 \mu 
\partial u/\partial z$ along the centerline of the channel. $\tau_{zz}$ 
vanishes 
where the flow is approximately parallel. As seen in Sec.\ 
\ref{tx:frontRearUzJumps}, $\tau_{zz}$ is negligible at small 
$Ca$, typically below $10^{-3}$.  

Furthermore, for a larger but still moderate $Ca$ number, it is observed in 
Fig.~\ref{fig:stressEvolution_Lam1_CaInf0.008_axis} that the pressure (red 
line) deviates from the linearly varying pressure, 
$p_{\text{linear}}$ (black line), of the unperturbed flow (without droplet) 
featured with a constant pressure gradient. The deviation is attributed to the 
non-parallel flow structure near the front and rear caps of the droplet (see 
Fig.\ 
\ref{fig:streamlines}), hence the pressure based on $p_{\text{linear}}$ need to 
be corrected by $\Delta p^{\text{NP}}=p-p_{\textsf{linear}}$. Typical values for the pressure corrections can be found in the 
Appendix \ref{tx:PressureCorr}. These corrections are particularly large
at large viscosity ratios for the region inside of the droplet. We did not succeed in providing a model to 
quantify this pressure correction.

Finally, in agreement with the results of Section \ref{tx:frontRearStressJump}, 
the jump in total stress or pressure at the rear of the droplet is smaller 
than the one at the front.

\subsection{Pressure distribution along the channel wall 
\label{tx:stressesEvolWall}}

The  pressure distribution on the channel wall is 
presented on 
Fig.\ \ref{fig:stressEvolutions} as well (continuous grey line). The influence 
of the interface mean curvature is clearly visible. The non-monotonic pressure 
at the wall close the droplet rear
results from the variation of the plane curvature in the dynamic meniscus 
region, where the interface oscillates (see also 
Fig.~\ref{fig:curvatureEvolution}).

\subsection{Droplet-induced total pressure drop along a channel 
\label{tx:totalPressureDrop}}

The prediction of the total pressure drop along a channel induced by 
the presence of a droplet flowing with a velocity $U_d$ is of paramount 
importance for the design of two-phase flow pipe networks 
\cite{baroud2010dynamics,ladosz2016pressure}. This allows for a coarse-grained quantification of 
the complicated local effects induced by the droplet. Droplets can thus be 
seen as punctual perturbations in the otherwise linear pressure 
evolution. In this section, we will show that it is possible to predict the 
total pressure drop induced by a droplet with the models proposed so 
far.

The total pressure drop can be defined as the difference
between the pressure in the outer phase ahead and behind the droplet, namely 
$\Delta p_{\textsf{tot}} = p_f^o-p_r^o$ \cite{kreutzer2005inertial}. It is given 
by
\begin{align}
	\Delta p_{\textsf{tot}} = &\Delta p_{o,r}^{\text{NP}} + \Delta p_r -\Delta p_{i,r}^{\text{NP}} \nonumber \\ 
	+&\frac{dp_i}{dz} L_d +\Delta p_{i,f}^{\text{NP}} - \Delta p_f - \Delta p_{o,f}^{\text{NP}} ,
\end{align}
where $\Delta p_{f,r}$ are given by the model for the pressure jumps at interfaces, equation \eqref{eq:DpfDprImproved}. The pressure gradient $dp_i/dz$ in the parallel region inside the droplet is given by 
Eq.\ \eqref{eq:dpdzParallelAxi} and Eq.~\eqref{eq:dpdzParallelPlanar} for the 
axisymmetric and planar geometries, respectively. 
Assuming the droplet of volume/area 
$\Omega$ (axisymmetric/planar geometry) as a composition of two hemispherical caps of radius 
$R-H_{\infty}$, with
$H_{\infty}$ given by Eq.\ \eqref{eq:HinfRational}, connected by a cylinder of the same radius, the droplet length 
$L_d$ can be approximated at first order for low $Ca$, \textit{i.e.} for $H_{\infty}/R\ll1$, as
\begin{equation}
	L_d = \frac{(R+2 H_{\infty})}{\pi}\frac{\Omega}{R^3}+\frac{2}{3}(R-H_{\infty})\\
\end{equation}
for the axisymmetric case and 
\begin{equation}
	L_d = \frac{(R+H_{\infty})}{2}\frac{\Omega}{R^2}+\frac{4-\pi}{2}(R-H_{\infty})
\end{equation}
for the planar case.

Equivalently, the total pressure drop can 
also be calculated using the models for the normal viscous stress jump, equation 
\eqref{eq:DUzfDUzr}, and the front and rear plane curvatures, equation 
\eqref{eq:KfKr}, yielding:
\begin{align}
	\Delta p_{\textsf{tot}} = &\Delta p_{o,r}^{\text{NP}} +\Delta \tau_{{zz}_r}+ \chi \gamma \kappa_r -\Delta p_{i,r}^{\text{NP}} \nonumber \\ 
	+&\frac{dp_i}{dz} L_d +\Delta p_{i,f}^{\text{NP}} -\Delta \tau_{{zz}_f} - \chi \gamma \kappa_f - \Delta p_{o,f}^{\text{NP}} ,
\end{align}
where $\chi=2$ for the axisymmetric configuration and $\chi = 1$ for the planar one.

If we neglect the non-parallel flows effects on the pressure, $\Delta p^{\text{NP}}$, the total pressure drop would then be:
\begin{equation}
	\Delta p_{\textsf{tot}} =  \Delta p_r +\frac{dp_i}{dz} L_d - \Delta p_f .
	\label{eq:DpSimplified}
\end{equation}

Neglecting the effects of the non-parallel flows would induce an 
error on the pressure drop, increasing with $Ca$. For a single droplet of volume 
$\Omega = 12.9$, the error of Eq.~\eqref{eq:DpSimplified} compared 
to the numerical results is less than $3 \%$ for $\lambda = 0$, but reaches 
$15 \%$ for $\lambda = 1$ and even $48 \%$ for $\lambda = 100$. It is thus 
important to include the corrections accounting for the 
non-parallel flow effects to predict the pressure drop accurately, 
especially when the viscosity ratios $\lambda 
\gtrapprox 1$. Numerical simulations are therefore crucial to achieve so.

\section{Conclusions \label{tx:conclusions}}
This paper generalizes the theory of a confined
bubble flowing in an axisymmetric or planar channel to droplets 
of non-vanishing viscosity ratios. Empirical models for the relevant quantities such as 
the uniform and minimal film thicknesses separating the wall and the droplet, the 
front and rear droplet plane curvatures, the total pressure drop in the channel and 
the droplet velocity are derived for the range of capillary numbers from 
$10^{-4}$ to $1$, and viscosity ratios ranging from the value $\lambda = 0$ for bubbles to highly-viscous droplets.
Following the work of Schwartz et al.~\cite{schwartz1986motion}, we extend
the low-capillary-number predictions obtained by the lubrication approach of 
Bretherton \cite{bretherton1961motion} for bubbles to viscous droplets.
Extensive accurate 
moving-mesh arbitrary Lagrangian-Eulerian (ALE) finite-element numerical 
simulations are performed for the viscosity-ratio range $\lambda \in [0-100]$ 
to build a numerical database, based on which we propose empirical models for 
the relevant quantities. The models are inspired by the low-$Ca$ theoretical 
asymptotes, but their validity range reaches large capillary numbers 
($Ca>10^{-3}$), where the lubrication approach no longer holds.

We have found that the uniform film thickness for $Ca<10^{-3}$ does
not differ significantly with that of a bubble as long as $\lambda<1$. For 
larger viscosity ratios, instead, the film thickness increases monotonically and 
saturates to a value $2^{2/3}$ times the Bretherton's scaling for bubbles when $\lambda > 
10^3$. The film thickness can be modeled by a rational function similar to 
that proposed by Aussillous and Qu\'er\'e \cite{aussillous2000quick} for 
bubbles, where the fitting coefficient $Q$ depends on the viscosity ratio. 
Furthermore, the uniform film thickness saturates at large capillary numbers to 
a value depending on $Q$. The minimum film 
thickness can be predicted analogously.
The velocity of a droplet 
can be unambiguously derived once the uniform film thickness is known. We have 
shown that considering the full expression of the droplet velocity is crucial as 
the asymptotic series for low $Ca$ has a very restricted range of validity for 
non-vanishing viscosity ratios.

Furthermore, we have found that the evolution
of the front and rear cap curvatures as a function of the capillary number differs 
from the one of the pressure jumps at the front and rear droplet interfaces. 
This is  due to the normal viscous stress jumps. The contribution of the 
jumps has been overlooked in the literature, though it has to be considered for 
$Ca>10^{-3}$. 
With all these models at hand, the pressure drop across a droplet can be 
computed, which will be valuable for engineering practices.

We also have shown that the flow patterns inside and 
outside of the droplet strongly depend on the capillary number and viscosity 
ratio. In particular, for $\lambda < 1/2$ ($\lambda < 2/3$) for the axisymmetric 
(planar) configuration, when the film thickness is larger than a critical value 
${H_{\infty}^{\star}}/{R}$, recirculating regions at the front and rear of the 
droplet disappear. Furthermore, the recirculation region in the outer phase 
detaches from the droplet's rear interface for large film thickness yet smaller than 
${H_{\infty}^{\star}}/{R}$, implying the disappearance of the inner 
recirculating region at the rear.

The considered problem in a planar 
configuration could be relevant for the study of a front propagation in a 
Hele-Shaw cell \cite{park1984two,reinelt1985penetration}, where the second-phase 
viscosity is non-vanishing. For instance, one could compute the amount of fluid 
left on the walls when a finger of immiscible fluid penetrates 
\cite{saffman1958penetration}. Furthermore, the problem in the planar 
configuration can be seen as a first step towards understanding the dynamics of 
pancake droplets in a Hele-Shaw cell \cite{huerre2015droplets,zhu2016pancake}.
Another possible outlook is the extension of the present theory to 
capillaries with polygonal cross sections, where the film between the droplet 
and the walls is not axisymmetric, but thick films known as \textit{gutters} 
develop in the capillary corners. Three-dimensional numerical simulations 
are then necessary to resolve this asymmetry. A force balance will determine the 
droplet velocity and an equivalent pressure drop model could be proposed for 
these geometries.

Despite the fact that this work was motivated by
the vast number of droplet-based microfluidic applications, the 
analytically derived equation \eqref{eq:SchwartzPlanar} serves as a 
generalization of the well known Landau-Levich-Derjaguin-Bretherton equation 
\cite{landaulevich,derjaguin1943,bretherton1961motion} when the second fluid has a non-negligible 
viscosity. This equation could therefore be adapted to predict the film 
thickness in coating problems with two immiscible liquids.

\clearpage

\appendix

\section{Derivation of the flow profiles in the thin-film region for the planar configuration \label{tx:flowProfilesPlanar}}

Consider an axial location in the thin-film region. The velocity profiles inside, $u_i$, and outside, $u_o$, of the droplet can be described by:
\begin{align}
	u_i(r) &= \frac{1}{2 \mu_i} \frac{d p_i}{dz} r^2 + A_i  r + B_i, \\
	u_o(r) &= \frac{1}{2 \mu_o} \frac{d p_o}{dz} r^2 + A_o  r + B_o,
\end{align}
where $p_i$ and $p_o$ are the inner, respectively outer, pressures, and $A_i$, $B_i$, $A_o$ and $B_o$ are real constants to be determined. Given the symmetry at $r=0$ of the inner velocity, $A_i=0$. The other constants are found by imposing the no-slip boundary condition at the channel walls $u(R) = -U_d$ in the droplet reference frame, the continuity of velocities at the interface located at $r=R-H$, $u_i(R-H) = u_o(R-H)$, and the continuity of tangential stresses at the interface
\begin{equation}
	\mu_i \frac{d u_i}{dz}\bigg\vert_{r=R-H} = \mu_o \frac{d u_o}{dz}\bigg\vert_{r=R-H}.
\end{equation}
Eventually one obtains:
\begin{align}
	A_o =& \frac{1}{\mu_o} \left(\frac{d p_i}{dz}-\frac{d p_o}{dz}\right)(R-H), \\
	B_i =& \frac{1}{2 \mu_i \mu_o} \left[ -(R - H)^2\frac{d p_i}{dz}  \mu_o + H \left(2 H \frac{d p_i}{dz} - H \frac{d p_o}{dz}\right. \right.  \nonumber \\
	&\left.\left.- 2 R \frac{d p_i}{dz} \right)\mu_i  \right] -U_d, \\
		B_o =& \frac{1}{2 \mu_o}\left[ \left( \frac{d p_o}{dz} - 2\frac{d p_i}{dz} \right) R^2- 2 H R \left( \frac{d p_o}{dz} - \frac{d p_i}{dz} \right) \right] -U_d.
\end{align}

\section{Derivation of the interface profile equation for the planar configuration \label{tx:derivationSchwartzPlanar}}

The flow rates at any axial location where the external film thickness is $H$ are:
\begin{align}
	q_i =& 2 \int_0^{R-H} u_i(r) dr \label{eq:QiFullPlanar} \\
	   = & \frac{1}{3\mu_i\mu_o}\left\{-(R-H)\left[3 H \left( H \left( \frac{d p_o}{dz} -2 \frac{d p_i}{dz} \right)+2\frac{d p_i}{dz} R\right)\mu_i \right. \right. \nonumber \\
	   +& \left.\left.2 \frac{d p_i}{dz} (R-H)^2\mu_o\right]\right\}-2U_d (R-H), \nonumber \\
	q_o =& 2 \int_{R-H}^{R} u_o(r) dr  \label{eq:QoFullPlanar} \\ 
	=& \frac{H^2}{3\mu_o}\left[H\left(3\frac{d p_i}{dz} -2\frac{d p_o}{dz}\right)-3\frac{d p_i}{dz} R\right]-2 U_d H . \nonumber
\end{align}
In the droplet reference frame, the flow rate of the inner phase has to vanish, 
$q_i =0$. Furthermore, in the region where the film is uniform 
(see Fig.~\ref{fig:curvatureEvolution}), $H=H_{\infty}$, the inner and outer 
pressure gradients have to be equal. Using these two conditions one can solve 
for the pressure gradient in the uniform film region
\begin{equation}
	\frac{d p}{dz} \bigg\vert_{r=R-H_{\infty}} \approx -\frac{6 \mu_i  U_d}{2 R^2-(4 -6\lambda) H_{\infty} R + (2-3 \lambda) H_{\infty}^2}
\end{equation}
and for the outer flow rate, where the limit $H_{\infty}/R \ll1 $ is considered:
\begin{align}
	q_o \approx & -2 H_{\infty} \left[\frac{2 R^2  - (4-3 \lambda) H_{\infty}  R +2(1- \lambda) H_{\infty}^2 }{
  2R^2  - (4 -6\lambda) H_{\infty}  R+(2-3 \lambda) H_{\infty}^2}\right]U_d \nonumber \\
  \approx &  -H_{\infty}  \left[\frac{2 R  - (4-3 \lambda) H_{\infty}  }{
  R  - (2 -3\lambda) H_{\infty} }\right]U_d
\end{align}
The pressure gradients in the dynamic meniscus regions are 
no longer equal and their difference is proportional to the deformation of the 
interface $r=R-H$. Under the assumption of a quasi-parallel flow, and neglecting 
the viscous contribution in view of the lubrication assumption, the Laplace law 
imposes:
\begin{equation}
	\frac{d p_i}{dz} - \frac{d p_o}{dz} = \gamma \frac{d^3 H}{d z^3}.
	\label{eq:LaplaceDerivation}
\end{equation}
Knowing $q_i$ and $q_o$, Eqs.\ \eqref{eq:QiFullPlanar}, \eqref{eq:QoFullPlanar} can be solved for the unknown pressure gradients $dp_i/dz$, $dp_o/dz$ as a function of $H$:
\begin{align}
 \frac{d p_i}{dz} \approx & \frac{ 3\lambda \left\{2H\left[H_{\infty}(3\lambda-2)+R\right]-3H_{\infty}\left[H_{\infty}(3\lambda-4)+2R\right]\right\}\mu_oU_d}{H(R-H)\left[H(3\lambda-4)+4R\right]\left[H_{\infty}(3\lambda-2)+R\right]}, \\
  \frac{d p_o}{dz} \approx & -6 \left\{\frac{R(H-H_{\infty})\left[3\lambda(H+H_{\infty})-2(H+2H_{\infty})\right]}{H^3\left[H(3\lambda-4)+4R\right]\left[H_{\infty}(3\lambda-2)+R\right]}\right. \nonumber \\
   & \left.+\frac{H H_{\infty}[H(2-3\lambda)^2+H_{\infty}(3(5-3\lambda)\lambda-4)]+2R^2(H-H_{\infty})}{H^3\left[H(3\lambda-4)+4R\right]\left[H_{\infty}(3\lambda-2)+R\right]}\right\}\mu_o U_d
\end{align}
 and substituted into Eq.\ \eqref{eq:LaplaceDerivation}. Following Bretherton \cite{bretherton1961motion}, the resulting equation can be put in an universal form by the substitutions $H = H_{\infty} \eta$ and $z = H_{\infty} (3 Ca)^{-1/3} \xi$. In the limit of $H_{\infty}/R\rightarrow 0$, the governing equation for the interface profile reads:
  \begin{equation}
 	\frac{d^3 \eta}{d\xi^3} =2\frac{\eta -1}{\eta^3}\left[ \frac{2+3 m (1+ \eta + 3 m \eta)}{(1+3m)(4+3m\eta)}\right].
	\label{eq:SchwartzPlanarDerivation}
 \end{equation}
 where 
\begin{equation}
	m = \lambda \frac{H_{\infty}}{R}
\end{equation}
is the rescaled viscosity ratio.

\section{Derivation of the droplet velocity model for the planar configuration \label{tx:velocityPlanar}}

The velocity profiles in the uniform film region have been derived in Appendix \ref{tx:flowProfilesPlanar}. In particular, the inner and outer volumetric fluxes are given by Eqs.\ \eqref{eq:QiFullPlanar} and \eqref{eq:QoFullPlanar}, respectively. At the location where $H=H_{\infty}$ the interface is flat and the pressure gradients are equal, $dp_i/dz=dp_o/dz=dp/dz$. Furthermore, mass conservation imposes that $q_o=  2R (U_{\infty}-U_d)$ and since we are in the reference frame of the droplet, $q_i=0$. The system of two equations can be solved for the pressure gradient
\begin{equation}
	\frac{d p}{dz} \bigg\vert_{r=R-H_{\infty}} =   \frac{-3 R U_{\infty} \mu_i}{(R-H_{\infty})^3+H_{\infty}(3R^2-3H_{\infty}R +H_{\infty}^2)\lambda} 
	\label{eq:dpdzParallelPlanar}
\end{equation}
 and the droplet velocity
\begin{equation}
	U_d=  \frac{R[2(R-H_{\infty})^2+3 H_{\infty}(2R-H_{\infty})\lambda]}{2(R-H_{\infty})^3+2H_{\infty}(3R^2-3H_{\infty}R+H_{\infty}^2)\lambda}U_{\infty} .
	\label{eq:UdPlanar}
\end{equation}
The relative velocity of the planar droplet reads
\begin{equation}
	\frac{U_d-U_{\infty}}{U_d} = \frac{\frac{H_{\infty}}{R} \left\{2-\frac{H_{\infty}}{R} \left[4+2 \frac{H_{\infty}}{R}(\lambda-1) -3 \lambda\right] \right\}}{2+\left(2-\frac{H_{\infty}}{R}\right)\frac{H_{\infty}}{R}(3\lambda -2)}.
\end{equation}

\section{Derivation of the critical uniform film thickness for the appearance of the recirculation regions \label{tx:derivationHstar}}

The velocity profile in the channel away from the droplet is given by Eq.~\eqref{eq:uInf} for the axisymmetric configuration and by
\begin{equation}
	u_{\infty}(y) = \frac{3}{2} U_{\infty}\left[1-\left(\frac{y}{R}\right)^2\right]-U_d
	\label{eq:uInfPlanar}
\end{equation}
for the planar one. The droplet velocity for the former case is given by Eq.~\eqref{eq:Ud}, whereas for the latter it is given by Eq.~\eqref{eq:UdPlanar}. With the use of Eqs.~\eqref{eq:Ud} and \eqref{eq:UdPlanar}, the velocity $u_{\infty}$ can be expressed as a function of $H_{\infty}/R$. The critical uniform film thickness for the appearance of recirculation regions, $\Hc$, is obtained by solving $u_{\infty}(0) = 0$, resulting in Eqs.~\eqref{eq:HcAxi} and \eqref{eq:HcPlanar} for the axisymmetric and planar configurations, respectively.

\section{Fitting laws for the model coefficients \label{tx:fittingLaw}}

The model coefficients $Q$ in Eq.~\eqref{eq:HinfRational}, $G$ in Eq.~\eqref{eq:HminRational}, $T_{f,r}$ and $Z_{f,r}$ in Eq.~\eqref{eq:KfKr} and $M_{f,r}$, $N_{f,r}$ and $O_{f,r}$ in Eq.~\eqref{eq:DUzfDUzr} can be well approximated by the rational function
\begin{equation}
	\frac{a_3 \lambda^3+a_2 \lambda^2+a_1 \lambda + a_0 }{\lambda^3+b_2 \lambda^2 +b_1 \lambda + b_0},
	\label{eq:fittingLaw}
\end{equation}
where the constants $a_i$ with $i=0,..,3$ and $b_j$ with $j=0,..,2$ are given in tables \ref{tab:cofficientsAxis} and \ref{tab:cofficientsPlanar} for the axisymmetric and planar geometries, respectively.

\begin{table}
\caption{Coefficients of the fitting law for the axisymmetric configuration.}
\begin{tabular}{lccccccc}
\hline\noalign{\smallskip}
 & $a_0$ & $a_1$ & $a_2$ & $a_3$    \\
\noalign{\smallskip}\hline\noalign{\smallskip}
$Q$ & $2.21$ & $111.25$ & $33.84$ & $1.37$    \\
$G$ & $130.37$ & $186.67$ & $-4.82$ & $1.30$   \\
$T_f$ & $3262.57$ & $1573.07$ & $7222.70$ & $9.90$    \\
$T_r$& $-12031.57$ & $-21476.98$ & $2820.73$ & $77.21$    \\
$Z_f$ & $3392.32$ & $-1773.73$ & $2984.79$ & $39.56$    \\
$Z_r$ & $-1842.14$ & $-14129.53$ & $26169.48$ & $160.45$    \\
$M_f$ & $-4850.40$ & $5797.90 $ & $-507.02$ & $1.22$   \\
$M_r$ & $-6.38$ & $18.59$ & $ -10.85$ & $-0.82$    \\
$N_f$ & $-5293.51$ & $14808.02$ & $-9344.15$ & $-126.15$    \\
$N_r$ & $-2.93$ & $-17.28$ & $18.94$ & $1.08$    \\
$O_f$ & $0.01$ & $-0.02$ & $0.08$ & $-0.11$   \\
$O_r$ & $32.38$ & $-429.86 $ & $638.07$ & $-5.84$    \\                            
\noalign{\smallskip}\hline
\end{tabular}
\label{tab:cofficientsAxis}
\end{table}
\begin{table}
\begin{tabular}{lccccccc}
\hline\noalign{\smallskip}
 &  $b_0$ & $b_1$ & $b_2$   \\
\noalign{\smallskip}\hline\noalign{\smallskip}
$Q$ &  $0.89$ & $44.86$ & $54.50$   \\
$G$ &  $58.41$ & $154.37$ & $10.56$   \\
$T_f$ &  $1197.26$ & $2006.27$ & $2855.96$   \\
$T_r$&  $25461.62$ & $11675.00$ & $16374.62$   \\
$Z_f$ &  $5249.41$ & $12649.53$ & $32757.19$   \\
$Z_r$ & $19514.41$ & $14458.49$ & $33771.45$   \\
$M_f$ & $2412.12$ & $ 2134.95 $ & $-222.42$   \\
$M_r$ & $ -2.68$ & $4.97$ & $2.93$   \\
$N_f$ &  $-3171.89$ & $-3079.05$ & $8185.38$   \\
$N_r$ & $1.68$ & $10.76$ & $5.71$   \\
$O_f$ &  $0.06$ & $-0.06 $ & $-0.64$   \\
$O_r$ &  $11.85$ & $-155.00$ & $ 338.88$   \\                            
\noalign{\smallskip}\hline
\end{tabular}
\label{tab:cofficientsAxis2}
\end{table}

\begin{table}
\caption{Coefficients of the fitting law for the planar configuration.}
\begin{tabular}{lccccccc}
\hline\noalign{\smallskip}
 & $a_0$ & $a_1$ & $a_2$ & $a_3$   \\
\noalign{\smallskip}\hline\noalign{\smallskip}
$Q$ & $98.76  $ & $      146.42 $ & $70.42$ & $ 1.45$ \\
$G$ &  $ 168.27  $ & $348.60  $ & $   26.76 $ & $  1.48$ \\
$T_f$  &  $0.35    $ & $ 1.17 $ & $  5.41  $ & $2.43$ \\
$T_r$ &  $ -130.18  $ & $  -298.84   $ & $-55.49  $ & $ -0.66 $ \\
$Z_f$ &  $ 1096.45 $ & $ 191.51  $ & $   395.61$ & $ -0.08 $ \\
$Z_r$&  $  -0.62   $ & $ -0.66 $ & $0.08 $ & $ -0.21 $ \\
$M_f$ & $-6.41$ & $17.26$ & $-11.54$ & $0.80$    \\
$M_r$ & $4.10$ & $-3.52$ & $0.17 $ & $0.04$   \\
$N_f$ & $ -6.12 $ & $17.95$ & $-11.90$ & $0.19$    \\
$N_r$ & $-50.40$ & $61.95$ & $-14.79$ & $0.51$    \\
$O_f$ & $4.52$ & $-2.80$ & $ -1.73$ & $0.89$   \\
$O_r$ & $2.10$ & $-6.97$ & $2.79$ & $ -0.01$   \\
\noalign{\smallskip}\hline
\end{tabular}
\label{tab:cofficientsPlanar}
\end{table}
\begin{table}
\begin{tabular}{lccccccc}
\hline\noalign{\smallskip}
 &  $b_0$ & $b_1$ & $b_2$   \\
\noalign{\smallskip}\hline\noalign{\smallskip}
$Q$ &  $ 45.04 $ & $ 89.61  $ & $ 77.03$\\
$G$ &   $ 88.60 $ & $ 264.97$ & $34.41$\\
$T_f$  & $  0.16  $ & $0.57$ & $ 3.02$\\
$T_r$ & $ 256.91 $ & $292.80  $ & $ 83.68$\\
$Z_f$ &   $  2690.72 $ & $ 6726.39 $ & $  2249.29$\\
$Z_r$&   $ 5.80   $ & $  -0.74 $ & $ 3.08$\\
$M_f$ &  $7.40$ & $ -2.88$ & $-12.76$   \\
$M_r$ &  $4.55$ & $7.67$ & $-5.25$   \\
$N_f$ &  $-9.06$ & $-6.06$ & $30.83$   \\
$N_r$ &  $84.57$ & $41.55 $ & $-24.20$   \\
$O_f$ &  $27.19 $ & $ -57.38$ & $21.87$   \\
$O_r$ &  $1.14$ & $-1.38$ & $-1.96$   \\
\noalign{\smallskip}\hline
\end{tabular}
\label{tab:cofficientsPlanar2}
\end{table}

\section{Additional results \label{tx:additionalResults}}

For seek of clarity, the results for $\lambda = 0$ and $100$ are shown in the appendix rather than in the main text, except for the normal viscous stresses jump, whose results for $\lambda = 0$ are presented in the main text as for $\lambda = 1$ the normal viscous stress jumps are small.

\begin{figure}[ht!]
\subfigure[\label{fig:FittingHinf_lambda0}]{
\includegraphics[width=0.45\textwidth]{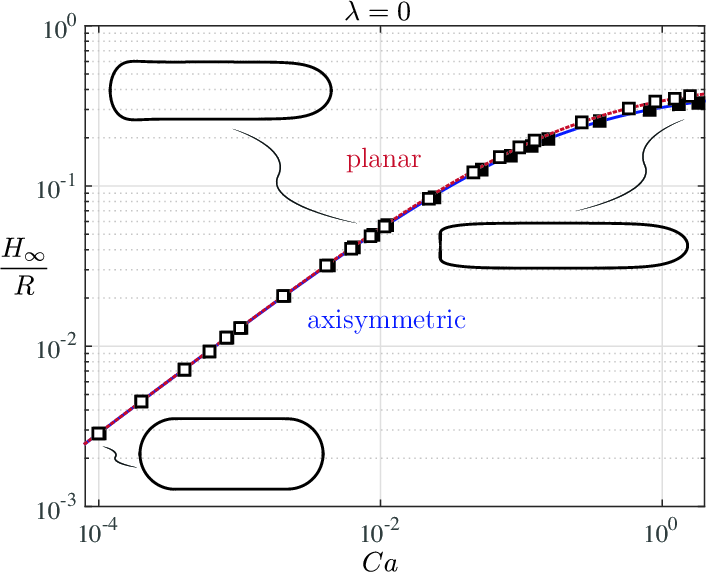}}
\subfigure[\label{fig:FittingHinf_lambda100}]{
\includegraphics[width=0.45\textwidth]{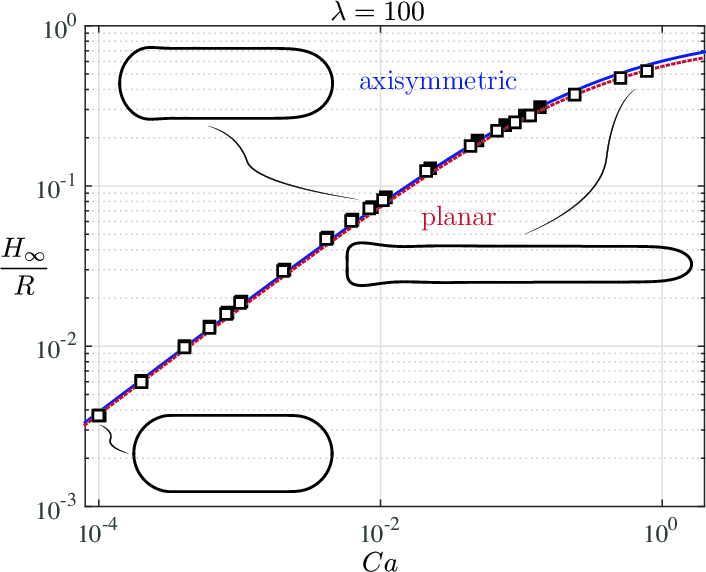}}
\caption{Uniform film thickness given by Eq.\ \eqref{eq:HinfRational} (lines) 
and FEM-ALE numerical results (symbols) as a function of the droplet capillary number 
for $\lambda = 0$ (a) and $100$ (b) and both axisymmetric (blue solid line, full 
symbols) and planar (dashed red line, empty symbols) geometries.
\label{fig:FittingHinf_lambda_appendix}}
\end{figure}
\begin{figure}[ht!]
\subfigure[\label{fig:FittingHmin_lambda0}]{
\includegraphics[width=0.45\textwidth]{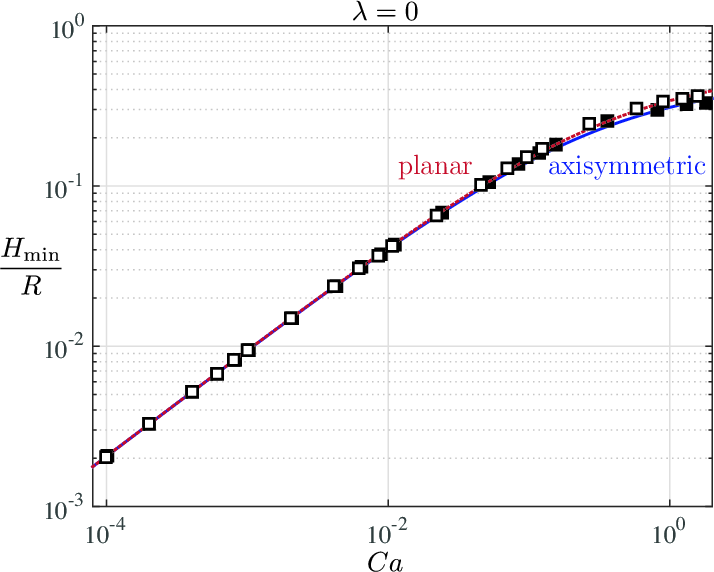}}
\subfigure[\label{fig:FittingHmin_lambda100}]{
\includegraphics[width=0.45\textwidth]{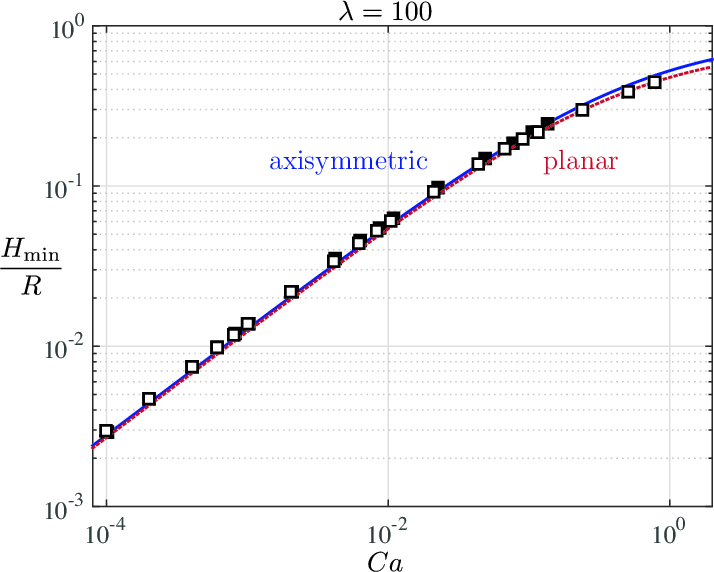}}
\caption{Minimum film thickness given by Eq.\ \eqref{eq:HminRational} (lines) 
and FEM-ALE numerical results (symbols) as a function of the droplet capillary 
number for $\lambda = 0$ (a) and $100$ (b) and both axisymmetric (blue 
solid line, full symbols) and planar (dashed red line, empty symbols) 
geometries. \label{fig:FittingHmin_lambda_appendix}}
\end{figure}

\begin{figure}[ht!]
\subfigure[\label{fig:FittingKf_lambda0}]{
\includegraphics[width=0.45\textwidth]{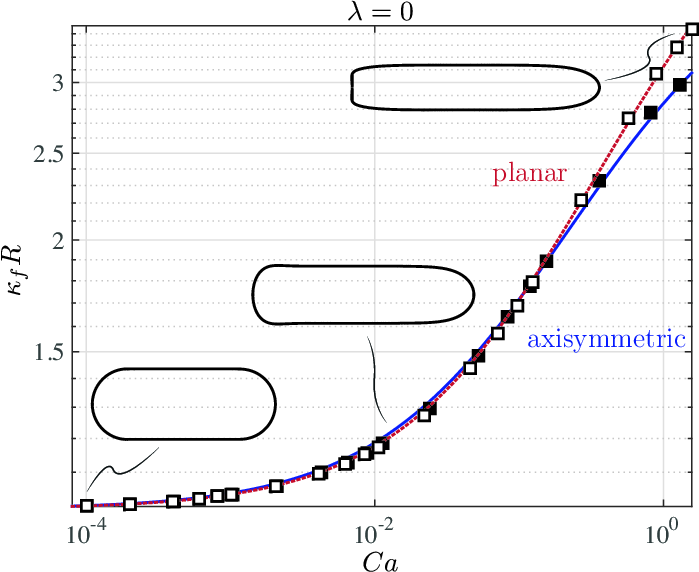}}
\subfigure[\label{fig:FittingKf_lambda100}]{
\includegraphics[width=0.45\textwidth]{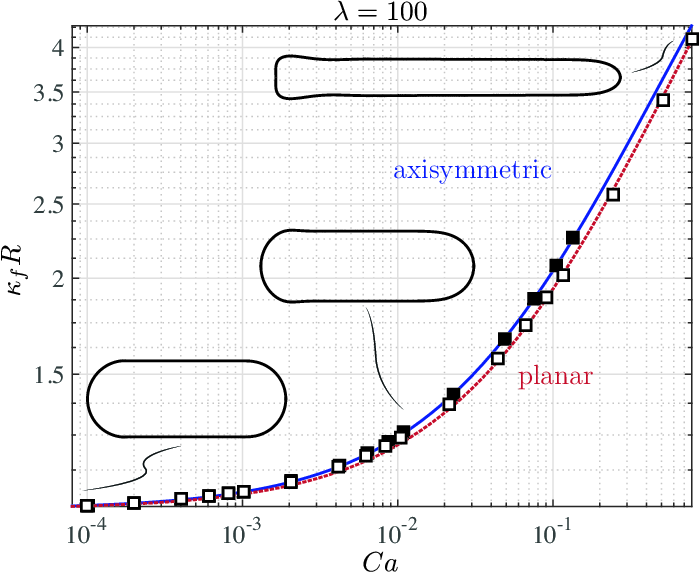}}
\caption{Curvature $\kappa_f$ of the front meniscus predicted by the model Eq.\ \eqref{eq:KfKr} (lines) 
and FEM-ALE data (symbols) versus $Ca$ for both axisymmetric (blue line, full
symbols) and planar (red dashed line, empty symbols) geometries, where the 
viscosity ratio $\lambda = 0$ (a) and $100$ (b). 
\label{fig:FittingKf_lambda_appendix}}
\end{figure}
\begin{figure}[ht!]
\subfigure[\label{fig:FittingKr_lambda0}]{
\includegraphics[width=0.45\textwidth]{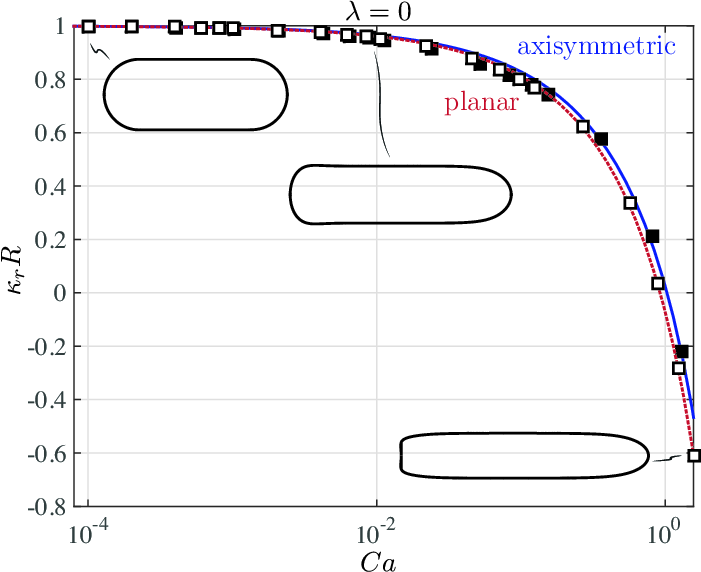}}
\subfigure[\label{fig:FittingKr_lambda100}]{
\includegraphics[width=0.45\textwidth]{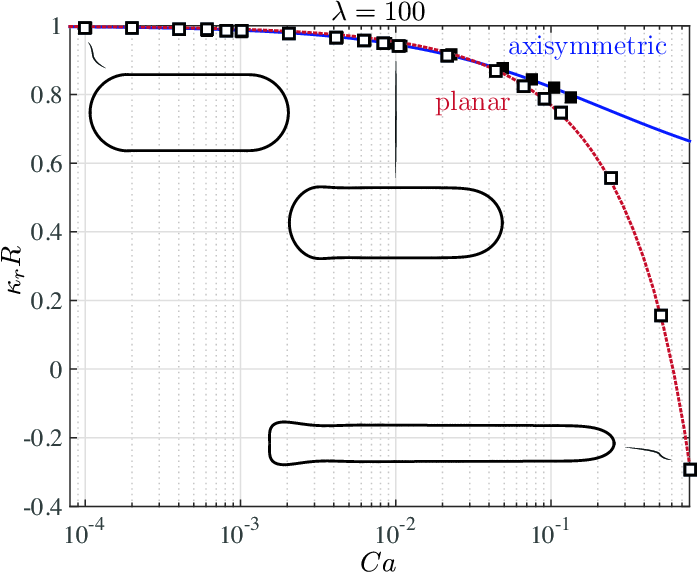}}
\caption{The rear counterpart $\kappa_r$ of Fig.~\ref{fig:FittingKf_lambda_appendix}.
\label{fig:FittingKr_lambda_appendix}}
\end{figure}

\begin{figure}[ht!]
\subfigure[\label{fig:FittingDpf_lambda1}]{
\includegraphics[width=0.45\textwidth]{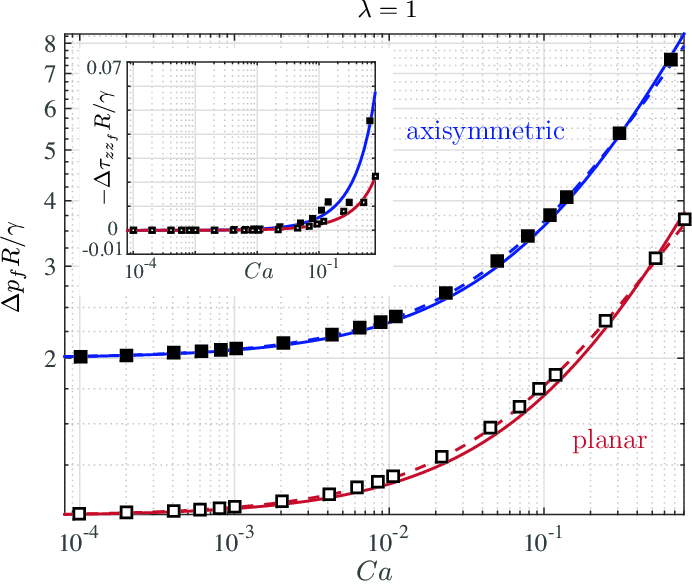}}
\subfigure[\label{fig:FittingDpf_lambda100}]{
\includegraphics[width=0.45\textwidth]{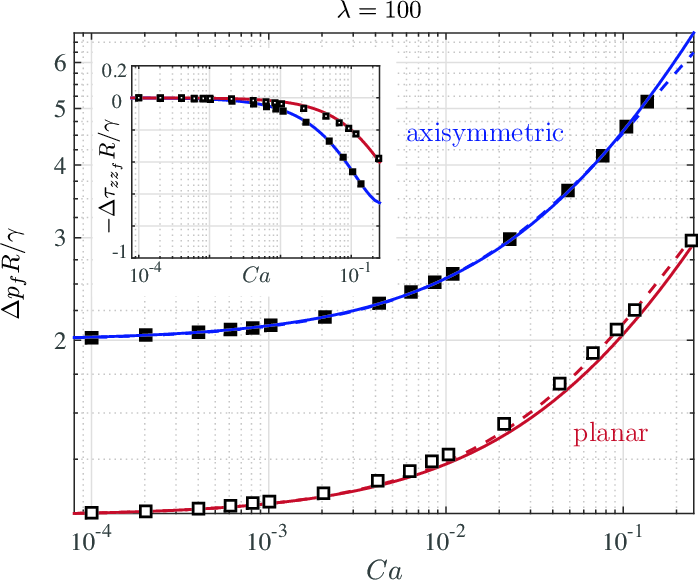}}
\caption{Front pressure jump $\Delta p_f$ given by Eq.\ \eqref{eq:DpfDpr} 
(solid lines) 
and front normal viscous stress jump $\Delta \tau_{{zz}_{f}}$  by Eq.\ 
\eqref{eq:DUzfDUzr} (inset, solid lines) and FEM-ALE data (symbols) versus 
$Ca$ for both axisymmetric (blue line, 
full symbols) and planar (red line, empty symbols) geometries, where the 
viscosity ratio $\lambda = 1$ (a) and $100$ (b). The dashed lines 
correspond to the improved pressure jump model Eq.\ \eqref{eq:DpfDprImproved}. 
Note the different scale in the insets. \label{fig:FittingDpf_lambda_appendix}}
\end{figure}
\begin{figure}[ht!]
\subfigure[\label{fig:FittingDpr_lambda1}]{
\includegraphics[width=0.45\textwidth]{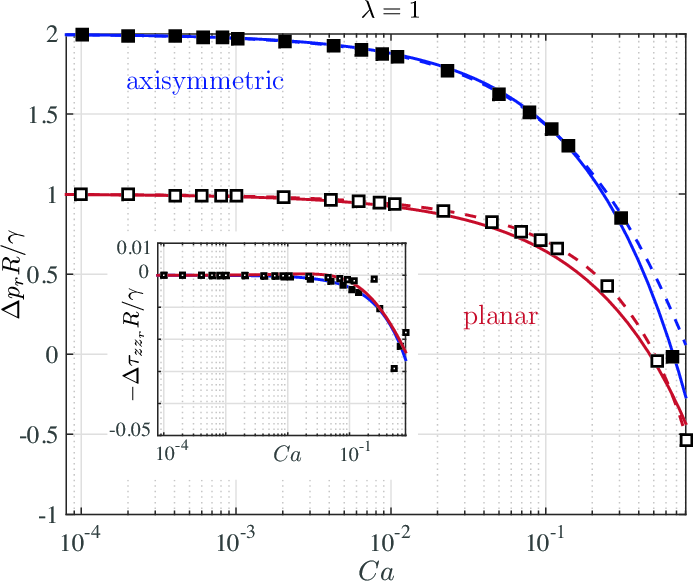}}
\subfigure[\label{fig:FittingDpr_lambda100}]{
\includegraphics[width=0.45\textwidth]{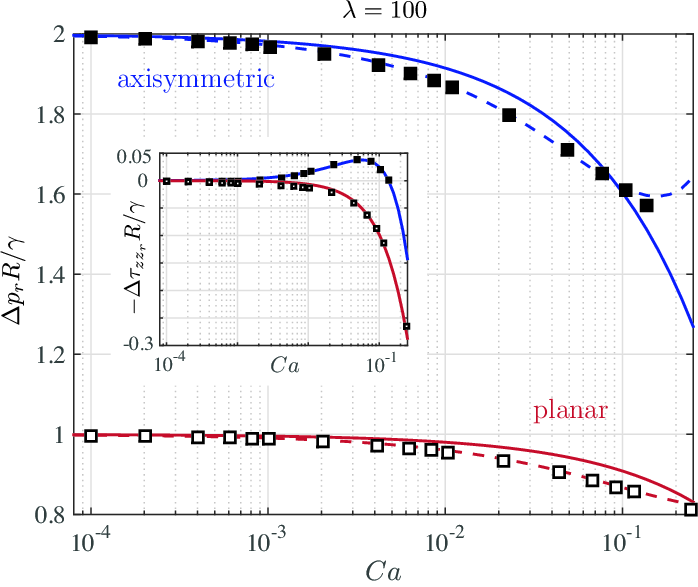}}
\caption{The rear counterpart, pressure jump $\Delta p_r$ 
and normal viscous stress jump $\Delta \tau_{{zz}_{r}}$, of 
Fig.~\ref{fig:FittingDpf_lambda_appendix}.
\label{fig:FittingDpr_lambda_appendix}}
\end{figure}

\section{Pressure corrections due to non-parallel flow \label{tx:PressureCorr}}

\begin{figure}[ht!]
\subfigure[\label{fig:stressCorr_rearOuter}]{
\includegraphics[width=0.45\textwidth]{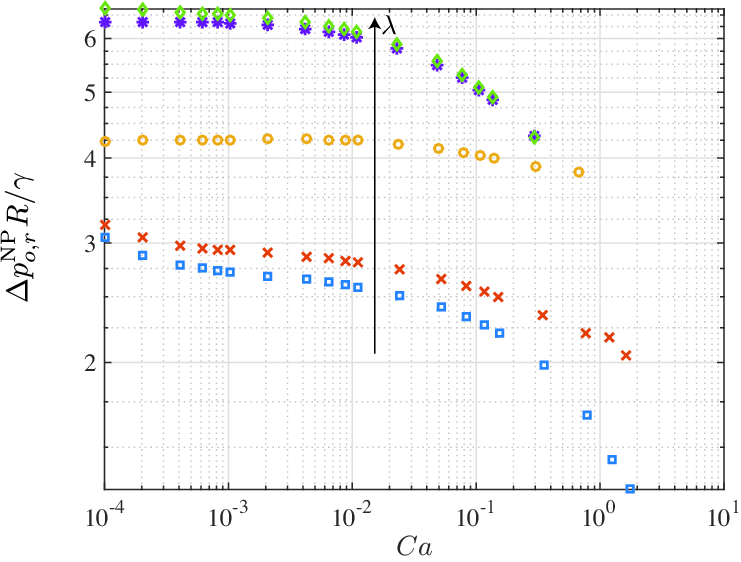}}
\subfigure[\label{fig:stressCorr_frontOuter}]{
\includegraphics[width=0.45\textwidth]{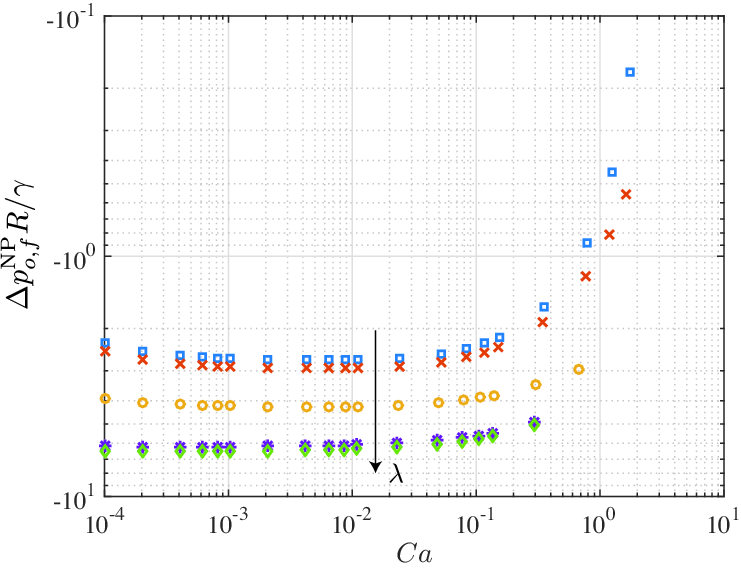}}
\caption{Pressure correction due to non-parallel flow effects at the rear (a) and front (b) outer sides of the interface for $\lambda = 0.04$ (blue squares), $0.12$ (red crosses), $1$ (yellow circles), $15$ (purple stars) and $50$ (green diamonds) for the axisymmetric configuration. The results are obtained from FEM-ALE numerical simulations. \label{fig:pCorrOuter}}
\end{figure}

\begin{figure}[ht!]
\subfigure[\label{fig:stressCorr_rearInner}]{
\includegraphics[width=0.45\textwidth]{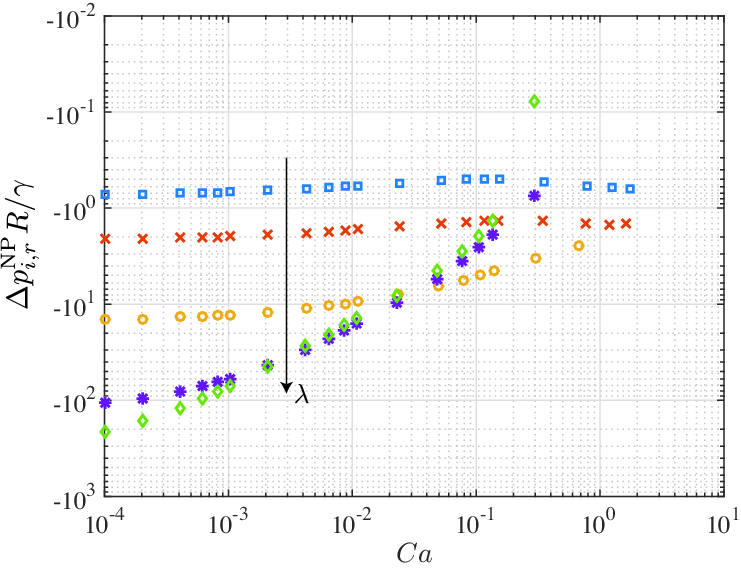}}
\subfigure[\label{fig:stressCorr_frontInner}]{
\includegraphics[width=0.45\textwidth]{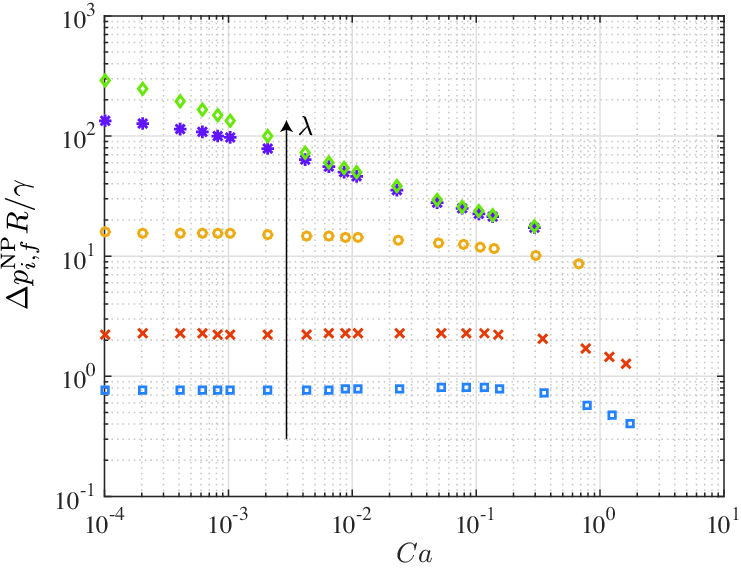}}
\caption{Pressure correction due to non-parallel flow effects at the rear (a) and front (b) inner sides of the interface for $\lambda = 0.04$ (blue squares), $0.12$ (red crosses), $1$ (yellow circles), $15$ (purple stars) and $50$ (green diamonds) for the axisymmetric configuration. The results are obtained from FEM-ALE numerical simulations.\label{fig:pCorrInner}}
\end{figure}

Some typical total stresses corrections at the outer and inner sides of the droplet interface as a function of $Ca$ and for different viscosity ratios $\lambda$ are shown in Fig.\ \ref{fig:pCorrOuter} and Fig.\ \ref{fig:pCorrInner}, respectively.

\begin{acknowledgements}

This work was funded by ERC Grant No.\ `SIMCOMICS 280117'. L.Z.\ gratefully acknowledges the VR International Postdoc 
Grant from Swedish Research Council `2015-06334' for financial support. 
The authors would like to acknowledge the valuable comments from the anonymous 
referees that helped to improve the manuscript.

\end{acknowledgements}

\clearpage

\bibliographystyle{spmpsci}      

%
%

\end{document}